\documentclass[11pt,onecolumn,english,amsart,preprintnumbers,amsmath,amssymb,floatfix,nofootinbib]{revtex4}
\usepackage{tikz,xcolor}
\usepackage[colorlinks = true,
linkcolor = red,
urlcolor  = blue,
citecolor = red,
anchorcolor = blue]{hyperref}

\definecolor{lime}{HTML}{A6CE39}
\DeclareRobustCommand{\orcidicon}{%
	\begin{tikzpicture}
	\draw[lime, fill=lime] (0,0)
	circle [radius=0.16]
	node[white] {{\fontfamily{qag}\selectfont \tiny ID}};
	\draw[white, fill=white] (-0.0625,0.095)
	circle [radius=0.007];
	\end{tikzpicture}
	\hspace{-2mm}
}

\foreach \x in {A, ..., Z}{%
	\expandafter\xdef\csname orcid\x\endcsname{\noexpand\href{https://orcid.org/\csname orcidauthor\x\endcsname}{\noexpand\orcidicon}}
}

\usepackage[toc,page]{appendix}
\usepackage{epsfig}
\usepackage{graphics}
\usepackage{latexsym}
\usepackage{amsmath}
\usepackage{amssymb}
\usepackage{rotating}
\usepackage{subfigure}
\usepackage{bm}
\usepackage{color}
\usepackage{ulem}
\usepackage{booktabs}

\begin{document}

%%%%%%%%%%%%%%%%%%%%%%%%%%%%%%

\title { Proton structure functions at low $\textit{x}$: the Fractal  distributions  }

\author{Samira Shoeibi Mohsenabadi$^{1,2}$\orcidA{}}

\author{Shahin Atashbar Tehrani$^{2}$\orcidA{}}

\author{Fatemeh Taghavi-Shahri$^{1}$\orcidC{}}
\email{taghavishahri@um.ac.ir (Corresponding author)}

\affiliation {
$^{(1)}$Department of Physics, Ferdowsi University of Mashhad, P.O.Box 1436, Mashhad, Iran  \\
$^{(2)}$School of Particles and Accelerators, Institute for Research in Fundamental Sciences (IPM), P.O.Box 19395-5531, Tehran,
Iran               }
\date{\today}

%
%%%%%%%%%%%%%%%%%%%%%%%%%%%%%%%%%%%%%%%%%%%%%%%%%%%%%%%%%%%
%
\begin{abstract}\label{abstract}
This paper presents a method for extracting the Parton Distribution Functions (PDFs) at small momentum fractions  $\textit{x}$ and at the next-to-leading order (NLO) accuracy in perturbative quantum chromodynamics. It turns out that the PDFs can be described by the "Fractal" or self-similar distributions at low $\textit{x}$ below $x<0.01$.
To this end, a simple parametrization for the unintegrated Parton Distribution Functions based on the "Fractal" approach is considered. These functions have self-similar behavior at low  $\textit{x}$ and  $k_t^2$ for sea quarks and have self-similar behavior at low $ x$ for gluon distribution. By integration from these TMDs, the initial input densities are obtained and the model's free parameters are then calculated using the experimental data released by the HERA experiment.
The small $\textit{x}$ experimental datasets on electron-proton ($e^-p$)  and positron-proton ($e^+p$) for natural current (NC) interactions in DIS processes at HERA  for the range of $1.2 < Q^2 < 500$ ($GeV^2$) and $x<0.01$ are included in this analysis.
The uncertainty estimations in the present analysis are carried out using the standard ``Hessian'' method.  Considering the overall value of $\chi^2/{\rm dof}$ and theory-to-data comparisons, the results indicate good agreements between the experimental datasets and the theoretical predictions.
A detailed comparison is also presented between the model's predictions for the relevant small-$\textit{x}$ observable and recent parameterizations for the PDFs.
\end{abstract}


\maketitle

\tableofcontents{}

%
%
%%%%%%%%%%%%%%%%%%%%%%%%%%%%%%%%%%%%%%%%%%%%%%%%%%%%%%%%%%%%%%%%%%%%%%%
\section{Introduction}\label{sec:introduction}
%%%%%%%%%%%%%%%%%%%%%%%%%%%%%%%%%%%%%%%%%%%%%%%%%%%%%%%%%%%%%%%%%%%%%%%
%
Quantum chromodynamics (QCD) is the fundamental theory of strong interaction. The nucleon structure and the physics of parton distribution functions (PDFs), especially in the context of deep-inelastic scattering (DIS), have been the subject of active theoretical and experimental research in the last decade.\\
Electrons and positrons deep inelastic scattering
off protons at HERA and center-of-mass energies of up to $\sqrt{s} \simeq 320\,$GeV allow us to explore the proton structure and quark-gluon dynamics through the perturbative QCD (pQCD) theory. H1 and ZEUS Collaborations, have explored a large phase space in  $(x, Q^2)$ and published data for the
cross-sections of neutral current (NC) interactions for $6 \cdot 10^{-7} \leq x \leq 0.65$ and $0.045 \leq Q^2 \leq 50000 $\, GeV$^2$
at values of the inelasticity, $y = Q^2/(sx)$,  between $0.005$ and $0.95$. They also published data on cross-sections of charged current (CC) interactions
for $1.3 \cdot 10^{-2} \leq x \leq 0.40$  and
$200 \leq Q^2 \leq 50000 $\,GeV$^2$
at values of inelasticity,  between $0.037$ and $0.76$ \cite{H1:2015ubc,H1:2010fzx}. \\
The combination of the data and the pQCD analysis led us to the proton structure. The proton structure is described
in terms of the parton distribution functions (PDFs), $q(x)$, which
are the probability of finding a parton, either gluon or quark,
with a fraction $\textit{x}$ of the proton's momentum and at a squared energy scale of $Q^2$. In the last decades, QCD analyses of deep inelastic scattering (DIS) data have been used to estimate parton densities  \cite{Ball:2017nwa,Dulat:2015mca,Jimenez-Delgado:2014twa,Butterworth:2015oua,Gao:2017yyd,Giuli:2019ptv}.
%**************************************************************
%**************************************************************
The experiments from DIS processes have shown that the number of these partons goes up at low $\textit{x}$, and falls at high $\textit{x}$ \cite{H1:2001ert}. Although at low $Q^2$, the three valence quarks become more dominant in the nucleon, at high $Q^2$ there are more and more sea quarks, the quark-antiquark pairs,  which carry a low momentum fraction x. In addition, the rising behavior of the proton structure function, $F_2(x, Q^2)$ with $Q^2$ at fixed small x, reveals a strongly increasing of the gluon density toward low x.
Knowledge of these densities at a much smaller value of $\textit{x}$ will be needed for any collider predictions. Therefore, it is crucial to know how parton distribution functions behave at low $\textit{x}$ \cite{xFitterDevelopersTeam:2018hym,Heidari:2019fio,Tokarev:2015qaa,Dremin:1992zc,Choudhury:2003yy,Choudhury:2005vy,Jahan:2011ig,Choudhury:2013ita,Jahan:2014ova,Jahan:2014sqa,Choudhury:2016fjy,Wilk:2013jsa,Boroun:2020fxg,Boroun:2018lpz,Deppman:2017igr}.\\
%**************************************************************
%***************************************************************
Here in this article, we attempt to explain the proton's internal sub-structure by describing its partonic distributions at low  $\textit{x}$ using the so-called  "self-similar" or "Fractal" distributions. \\
The topic of Fractal structures in hadronic systems has gained increasing attention in recent years from both the theoretical and experimental communities. Models based on Fractal geometries provide alternative perspectives on non-perturbative QCD phenomena like multi-scale correlations and emergent behaviors observed at high energies \cite{Deppman:2017igr,Deppman:2020jzl}.\\
Fractal parton studies shed new light on small-x physics and many-body dynamics in QCD. There are several important reasons to study Fractal models of parton distributions. Fractality may provide insights into complex multi-scale correlations arising from non-perturbative many-body QCD dynamics. It is directly relevant for probing small x physics where nonlinear QCD effects dominate, relating to geometric scaling and saturation phenomena. Developing self-similar Fractal structures could signal the onset of collective parton behavior. Additionally, Fractal models connect to concepts like critical exponents and phase transitions, which may be pertinent to QCD issues like deconfinement. Studying Fractal distributions also has implications for modeling tasks like PDF parametrization and event generation over broad energy ranges. Precisely testing predictions from Fractal models against data furnishes an understanding of how coherence emerges in non-perturbative QCD.\\ Furthermore, several experiments are currently pursuing lines of investigation to further explore the possibility of Fractal structures in parton distributions. Ongoing runs at LHCb seek to precisely map out small-x behavior and search for signatures that could validate Fractal scaling laws or the existence of phase transitions through high-precision measurements. Additional studies at the LHC, such as those analyzing proton-proton and proton-nucleus collisions to probe initial state fluctuations through quantities like multiplicity, flow, and jet substructure, may provide further insights. Plans for a future electron-ion Collider include using high-luminosity lepton-nucleus collisions to precisely determine nuclear PDFs and investigate spatial correlations in a way that could reveal Fractal patterns. The LHC heavy-ion program aims to understand how properties of the quark-gluon plasma relate to predictions stemming from Fractal initial state models. Projects like the Forward Physics experiment at the LHC also target uncharted small-x territories that may hold clues. Potential future hadron and lepton colliders could directly search for evidence of Fractal behavior to even smaller momentum fractions. Together, these ongoing and upcoming experimental investigations seek to accumulate sensitive data to better characterize the possible Fractal nature of QCD interactions \cite{Li:2022cwt,Zborovsky:2021evo,Megias:2022ksp,CMS:2021lab,ALICE:2013wgn,Fazio:2017zuj,Bluhm:2020mpc,Levin:2011hr}.	
\\
The paper is organized as follows. In Sec.II, we study the methods. In this section, we have a short review of the proton structure at low $\textit{x}$. We will describe the Fractal distributions and the initial input PDFs to solve the DGLAP evolution equations. We then discuss the experimental observable and the minimization processes. Then, we bring our results in section III. Finally, our summary and conclusions are presented in Sec. IV.
%--------------------------------------------------

%
%%%%%%%%%%%%%%%%%%%%%%%%%%%%%%%%%%%%%%%%%%%%%%%%%%%%%%%%%%%%%%%%%%%%%%%
\section{Methods}\label{sec:lowx}
%%%%%%%%%%%%%%%%%%%%%%%%%%%%%%%%%%%%%%%%%%%%%%%%%%%%%%%%%%%%%%%%%

%%%%%%%%%%%%%%%%%%%%%%%%%%%%%%%%%%%%%%%%%%%%%%%%%%%%%%%%%%%%%%%%%%%%%%%
\subsection{A short review of the proton structure at low $x$}\label{sec:lowx}
%%%%%%%%%%%%%%%%%%%%%%%%%%%%%%%%%%%%%%%%%%%%%%%%%%%%%%%%%%%%%%%%%
One interesting result from HERA is the rise of the proton structure function, $F_2(x, Q^2)$,  with decreasing the value of  $\textit{x}$ at fixed $Q^2$. Therefore, understanding the physics behind this behavior and investigating the proton structure function and the parton distribution functions at low $\textit{x}$ region is incredibly  important.\\
To study the proton internal structure at low $\textit{x}$, one needs to investigate the  reduced cross-section data in $e^{\pm}p$ DIS processes at low x region.  For unpolarized $e^{\pm}p$ scattering,
the reduced cross- section at low $Q^2$, i.e. $Q^2 \ll M_Z^2$,
can be written as \cite{H1:2015ubc,H1:2010fzx},
\begin{equation}\label{Eq:eq1}
\sigma_{r,NC} = F_2(x,Q^2)  - \frac{y^2}{Y_+} F_{\rm L}(x,Q^2)~.
\end{equation}
The kinematical variables $x$, $Q^2$ and $y$ are defined as
\begin{equation}
	Q^2=-q^2,\:x=\frac{Q^2}{2(P.q)},\:y=\frac{(P.q)}{P.k},
\end{equation}
P, k, and q denote the four-momentum of the incoming proton,  incoming lepton, and exchanged boson respectively, and  $Y_+=1+(1-y)^2$. Note that for the values of $y$ larger than approximately 0.5, the 
longitudinal structure function, ${F_{\rm L}}$ contribution  is  significant .\\
 The proton structure function,   $F_2 (x,Q^2)$  in  Eq. ~(\ref{Eq:eq1})  can be expanded as a function of the parton distribution functions, PDFs. Then we have \cite{Gluck:1989ze}
\begin{equation}
F_2(x,Q^2)= \sum\limits_q e^2_q\bigl\{ (1+\frac{\alpha_s(Q^2)}{4\pi}C_q) \otimes(q+ \bar q)(x,Q^2) +\frac{1}{3} \frac{\alpha_s(Q^2)}{4\pi}( C_g\otimes g)(x,Q^2)\bigr\}~,
\end{equation}
where  $e_q$ denotes the electric charge of  quarks and $ C_q$ and $C_g$ are the quarks and gluon Wilson coefficients respectively, and  convolution integral is defined as
\begin{equation}
(P\otimes q)(x,Q^2)=\int\limits^1_x \frac{dy}{y} P(\frac{x}{y}) q(y,Q^2).
\end{equation}
The $Q^2$ dependence of the parton distribution functions can be evaluated by the well-known  Dokshitzer-Gribov-Lipatov-Altarelli-Parisi  (DGLAP)  evolution equations using the suitable initial inputs \cite{Gribov:1972ri,Lipatov:1974qm,Altarelli:1977zs,Dokshitzer:1977sg}
%******************************************
\begin{eqnarray}\label{eq:dglap}
\frac{d q(x,Q^2)}{d ln( Q^2)}& = \frac{\alpha_s}{2\pi}&  \int_{x}^{1} \frac{dz}{z} (P_{qq}(\frac{x}{z})q(z,Q^2)+P_{qg}(\frac{x}{z})g(z,Q^2)),\\
\frac{d g(x,Q^2)}{d ln( Q^2)}& = \frac{\alpha_s}{2\pi}&  \int_{x}^{1} \frac{dz}{z} (\sum_{q,\bar q} P_{gq}(\frac{x}{z})q(z,Q^2)+P_{gg}(\frac{x}{z})g(z,Q^2)),
\end{eqnarray}
where the splitting functions $P_{ij}$  at  LO and NLO approximations are given in \cite{Altarelli:1977zs,Furmanski:1980cm,Furmanski:1981cw}.
Solution of the DGLAP evolution equations using the suitable initial inputs led us to  the quark and gluon distribution functions inside proton.
\\
To describe the low $\textit{x}$ behavior of the PDFs, there are two commonly used evolution equations: the DGLAP and the BFKL evolution equations \cite{Kuraev:1977fs,Balitsky:1978ic}. However, both of them are applicable to describe the internal structure of the protons in different phase-space regions. The DGLAP evolution equations study the $Q^2$ evolution of the parton distribution functions. It is shown that the DGLAP evolution equations can be solved in double leading log approximation and the proton structure shows to rise approximately as a power of $\textit{x}$ toward low $\textit{x}$. It contains the $\alpha _s ln ( \frac {Q^2}{Q_0^2})$ expansion terms \cite{Ermolaev:1999jx,Dokshitzer:1991wu}. Furthermore, the BFKL equation investigates the x evolution of the PDFs. In this equation, the terms in the form of  $\alpha _s ln ( \frac {1}{x})$ are considered.  The experimental data released by HERA can well described by the DGLAP evolution equations and also with  inclusion of the low x BFKL terms \cite{Ball:2017otu}.
These  two different approaches imply that the proton structure function at low $\textit{x}$ can be parametrized  as
\begin{equation}
F_2(x, Q^2) = c(Q^2) x^{- \lambda (Q^2)}~.
\end{equation}
where two functions $c(Q^2)$ and $\lambda (Q^2)$ can be determined by using experimental data for $F_2(x, Q^2)$ at low $\textit{x}$.\\
Theoretically, the rising behavior of proton structure function at low $\textit{x}$ can also be described by the Regge theory too \cite{Regge:1959mz,Donnachie:1992ny}. In this theory the total cross-section of hadron-hadron and photon- hadron scattering processes, is determined by the pomeron intercept $\alpha_P=1+\Delta$ , and is given by $ \sigma _{\gamma (h)p}^{total} \sim \nu ^\Delta$.
 This behavior is also valid for the virtual photon in the region of  $\textit{x} \ll 1$, leading to the well-known behavior of $F_2 \sim  x^{- \Delta}$.\\
The rise behavior of the proton structure function at low $\textit{x}$ may also be related to the so-called saturation effect that refers to the gluon recombination in the nucleon \cite{Rezaei:2018tie}.
Saturation phenomenology and the physics related to it, are expected to be relevant for any collider process that involves small-x partons for a broad range of observables and various collision systems, including collisions at RHIC and also at HERA.\\
In the next, we will describe the Fractal distributions that we used to determine the initial parton distribution functions at the low $\textit{x}$ region to solve the DGLAP evolution equations.

%-PLOTS---------------------------------
\begin{figure}[htb]
	\begin{center}
		\vspace{0.50cm}
		\resizebox{.450\textwidth}{!}{\includegraphics{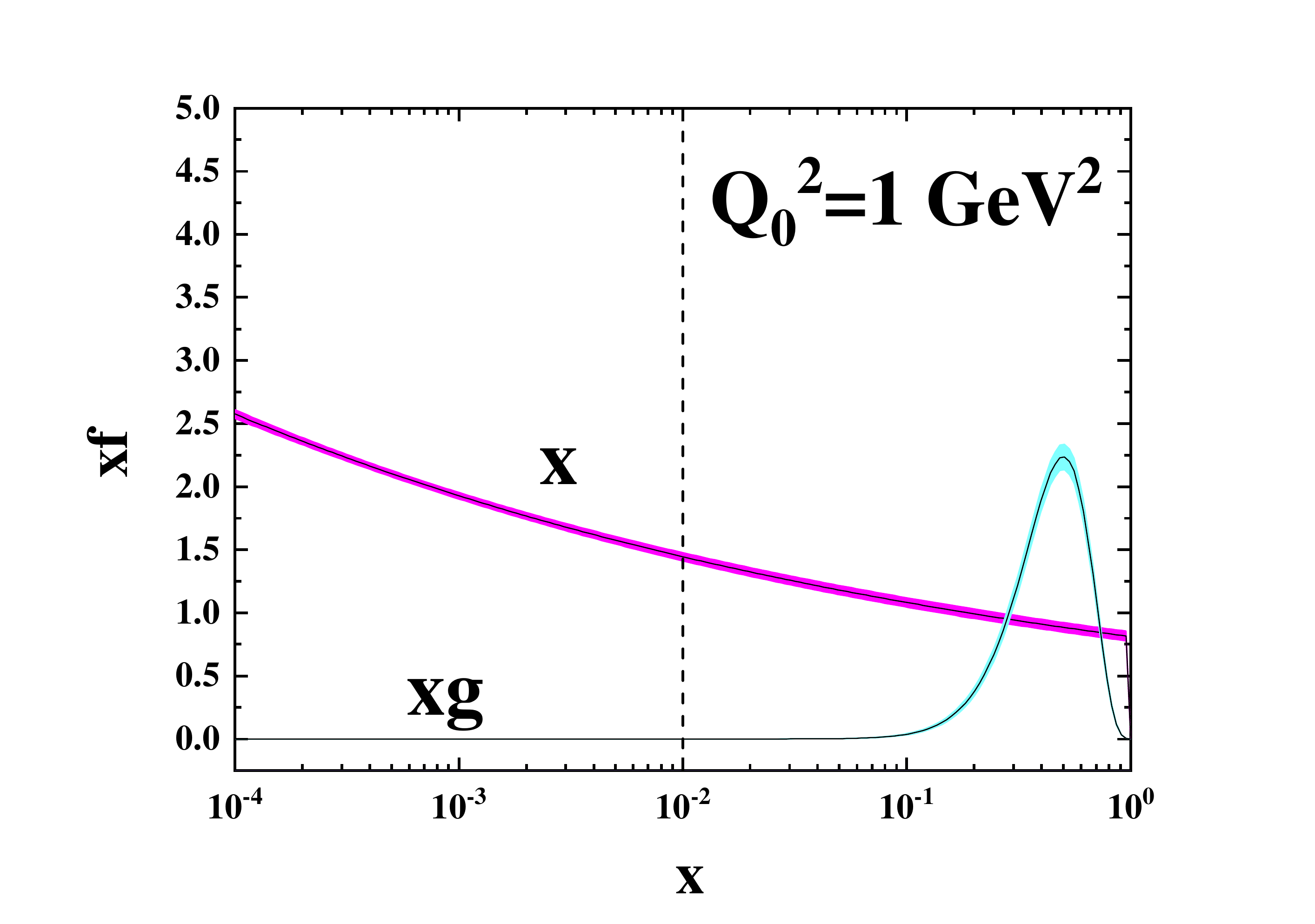}}
		\resizebox{.45\textwidth}{!}{\includegraphics{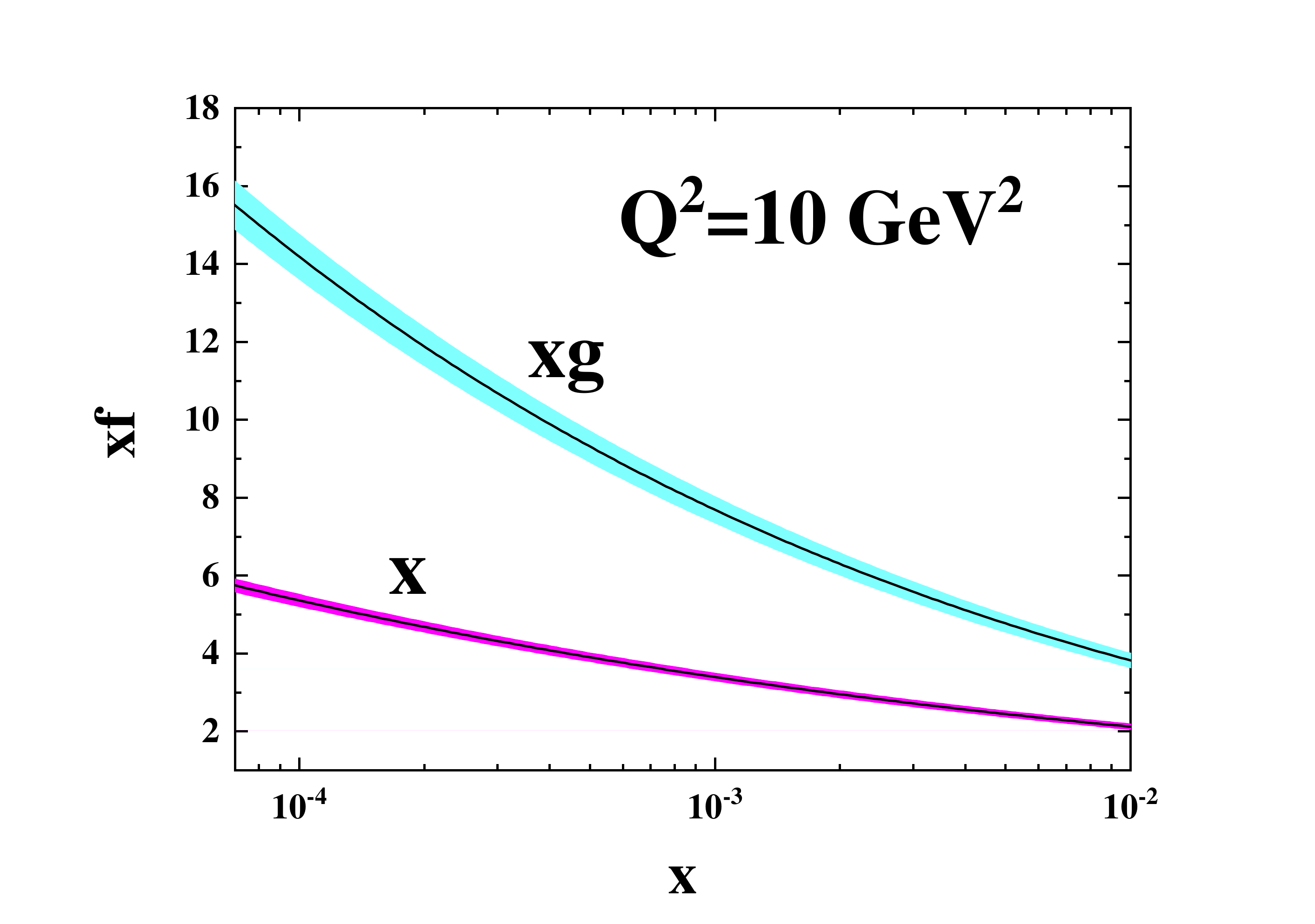}}
		\caption{{\small  (left) The parton distribution functions at $Q_0^2=1 GeV^2$. (Right) The same but for $Q^2=10 GeV^2$ and in the low x region below $x< 0.01$. Note that the Fractal PDFs are valid only for $x<0.01$.}} \label{fig:inputpdfs}
	\end{center}
\end{figure}

\begin{figure}[htb]
	\begin{center}
		%\vspace{0.5cm}
		\resizebox{0.45\textwidth}{!}{\includegraphics{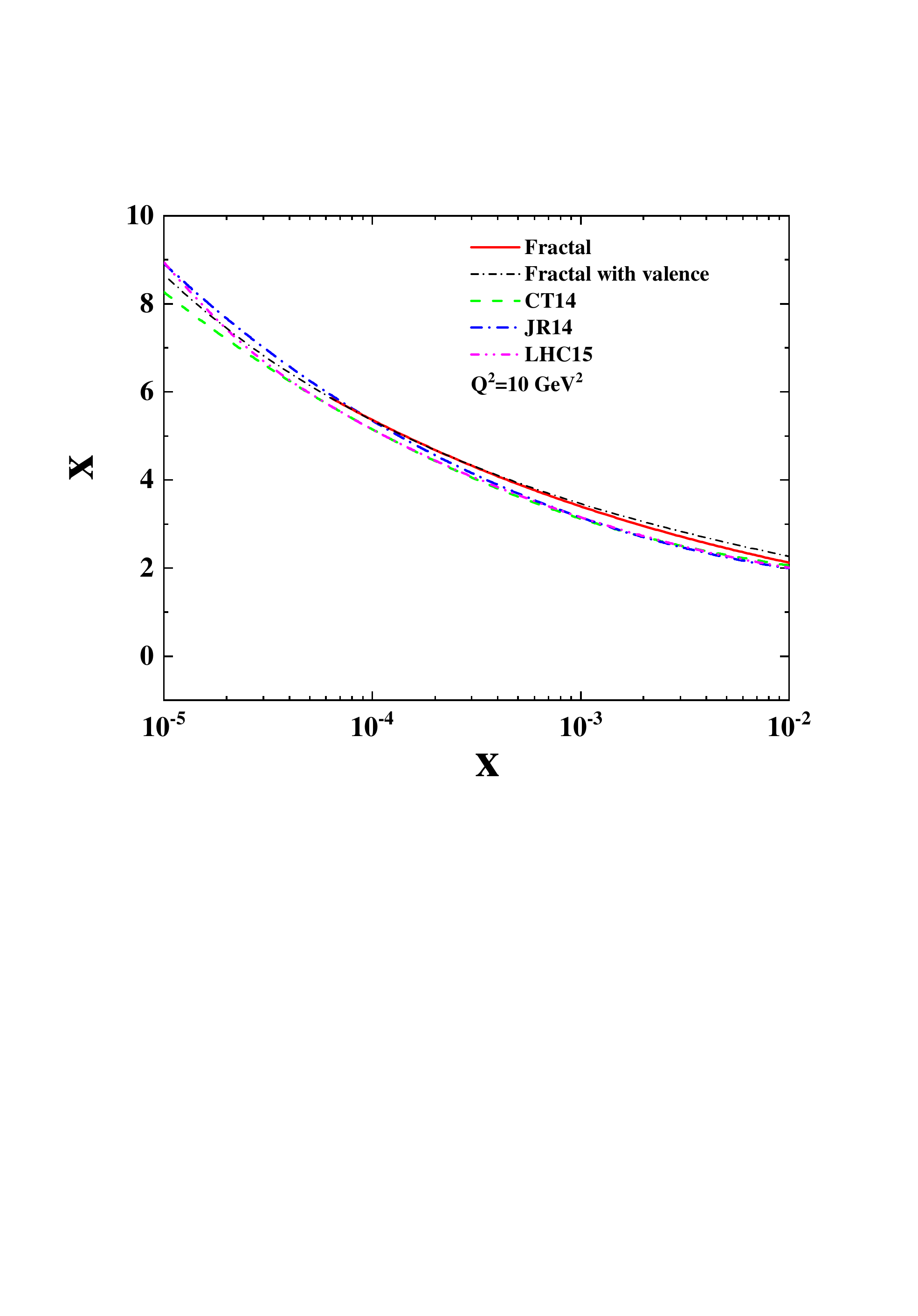}}
		\resizebox{0.45\textwidth}{!}{\includegraphics{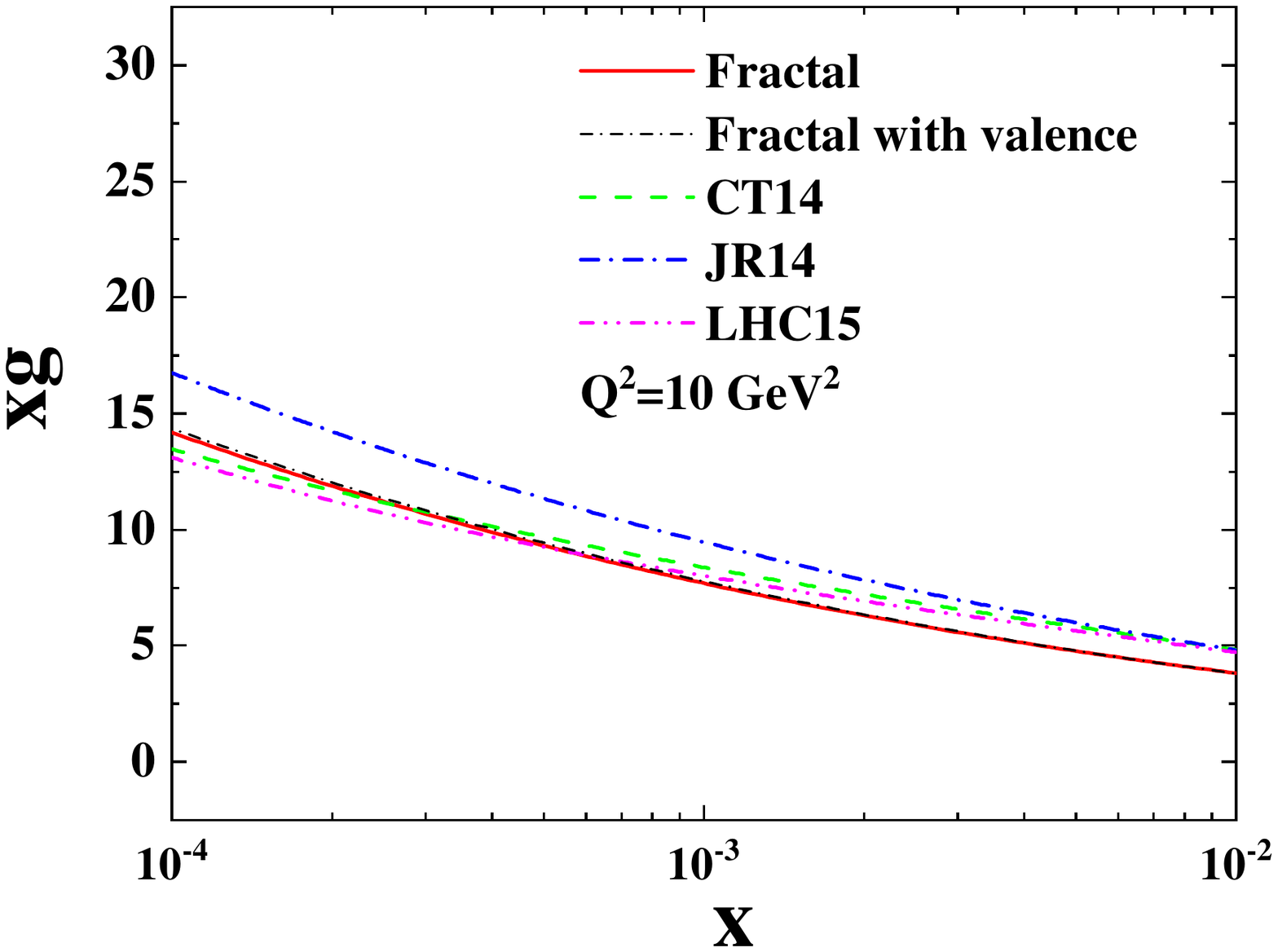}}
		\vspace{-3.5cm}
		\caption{{\small The parton distribution functions at low x and comparison with those from CT14, JR14, and LHC15 parameterizations at $Q^2=10 GeV^2$ \cite{Dulat:2015mca,Jimenez-Delgado:2014twa,Butterworth:2015oua}.} \label{fig:pdfsq10}}
	\end{center}
\end{figure}

\begin{figure}[htb]
	\begin{center}
		\vspace{0.50cm}
		\resizebox{.450\textwidth}{!}{\includegraphics{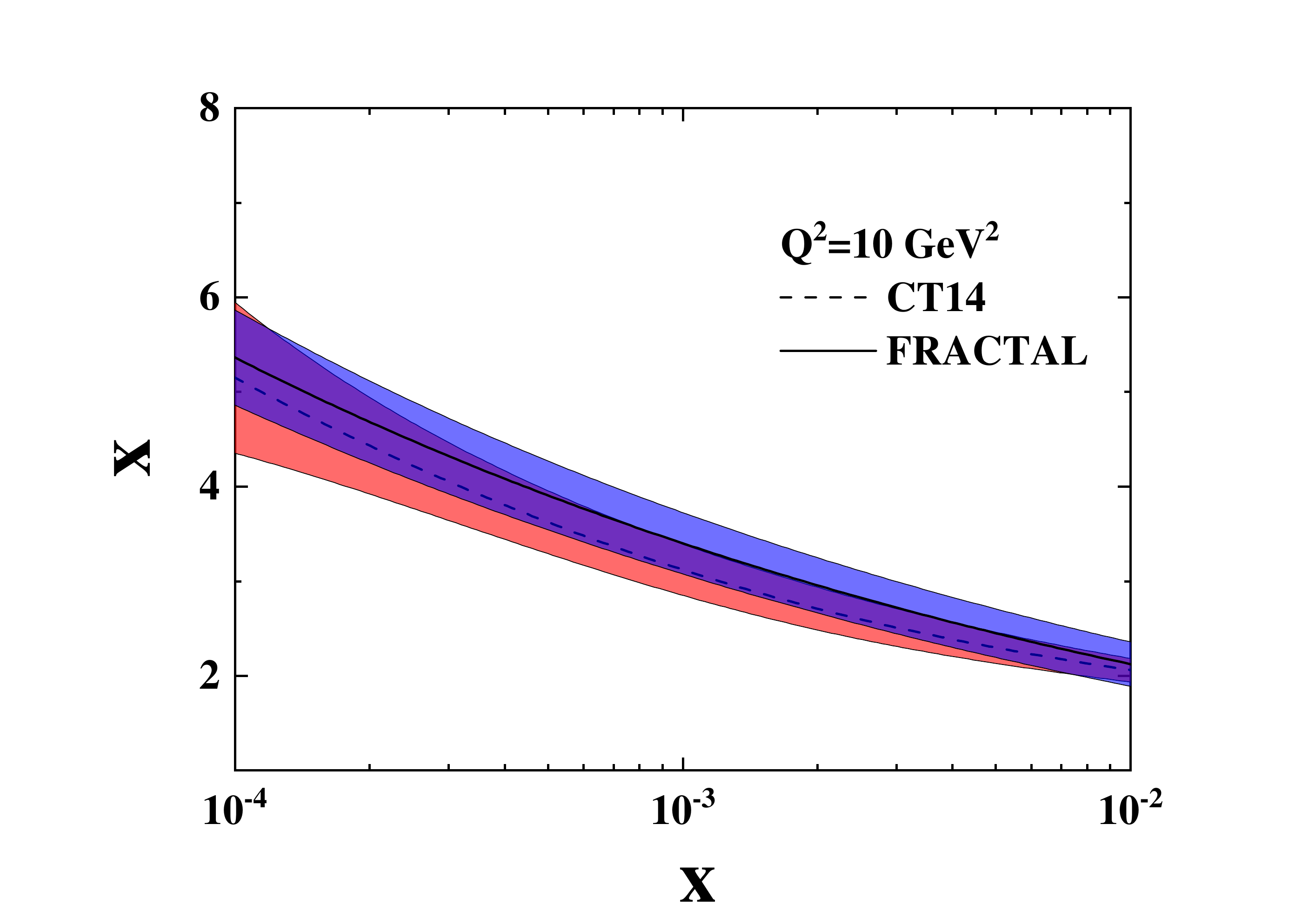}}
		\resizebox{.450\textwidth}{!}{\includegraphics{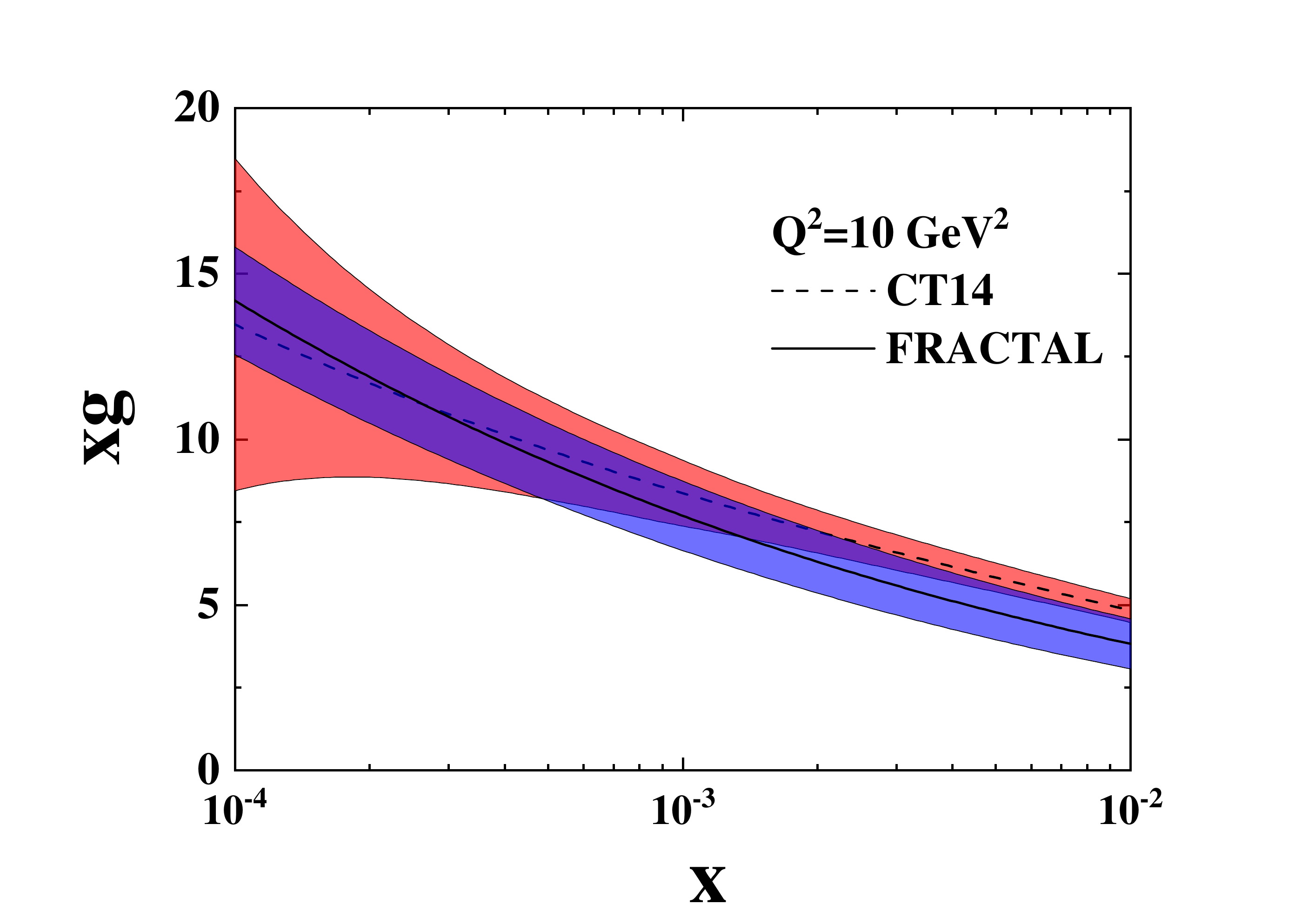}}
		\caption{{\small The parton distribution functions  with uncertainties band at low x in comparison with those from CT14 parameterization at $Q^2=10 GeV^2$ \cite{Dulat:2015mca}. } \label{fig:pdfsq10ct}}
	\end{center}
\end{figure}

%%%%%%%%%%%%%%%%%%%%%%%%%%%%%%%%%%%%%%%%%%%%%%%%%%%%%%%%%%%%%%%%%%%%%%%
\subsection{ The Fractal distributions and the initial input PDFs  } \label{sec:parametrization}
%%%%%%%%%%%%%%%%%%%%%%%%%%%%%%%%%%%%%%%%%%%%%%%%%%%%%%%%%%%%%%%%%%%%%%%
%
%

In this section, we will describe the phenomenological parametrizations as well as the assumptions we used in our analysis. Here, we used the "Fractal" distributions to describe the parton distribution functions inside the proton at low $\textit{x}$.
We show that the sea quark distribution functions have "Fractal" or self-similar behavior with a fixed exponent at $x<0.01$ and also at low $k_t^2$ whereas the gluon density has this Fractal behavior at low $\textit{x}$. After a short introduction to the "Fractal" and Fractal distributions, we try to describe the TMDs and the PDFs inside the proton with a mono Fractal approach that uses the fixed exponents for the initial parton densities.\\
The Fractal based description of the DIS processes and the proton structure functions were initially introduced in \cite{Choudhury:2003yy,Jahan:2014sqa,Choudhury:2016fjy,Lastovicka:2002hw,Lastovicka:2004mq,Hwa:1989vn,Florkowski:1990ba} to study the proton structure at low $\textit{x}$. In these researches, the quark distribution functions are calculated at LO (Leading- Order) approximation without the inclusion of the gluon distribution functions. Here we did the same   at NLO (Next- to- Leading- Order) approximation. Moreover, here the gluon distribution function plays a role at low $\textit{x}$.We will show that incorporating
	the gluon distribution in the low x ($Q^2$) region leads to a better understanding
	of the low x ($Q^2$) region.This model is capable of ”predicting” the
	experimental data related to the reduced cross-section, and heavy quark structure functions accurately. By incorporating the gluon distribution function in
	this region, we can provide a more accurate description of the physics occurring
	here.\\
The Fractal concept which is initially introduced and expanded by Mandelbrot \cite{Mandelbrot}, is an important tool used to describe and model complex systems in nature.
Fractals are never-ending patterns that have several properties such as self-similarity and iterative formation. In mathematical language, similarity means that when the size is varied, the form would not be changed.\\
The feature that plays a key role in our analysis is the Fractional or Fractal dimension. In order to consider this concept, first we should find out what is the relation between the definition of the dimension and self-similarity. It has been shown that the concept of geometric dimension can be connected to the inherent self-similarity of Fractal objects. Specifically, this relationship known as the Fractal dimension formula
	\begin{equation}\label{Eq:Fractal-Dimension1}
		D = \frac{\log(N)}{\log(r)},
	\end{equation}
		where N is the number of self-similar pieces created by segmenting the original shape into r equal parts, with r representing the magnification factor. When applying this to well-known Fractals such as the Sierpinski gasket, the resulting Fractal dimension value of approximately 1.585 is obtained, demonstrating it is a non-integer dimension consistent with Fractal geometry. As described by Mandelbrot, this Fractal dimension expression provides a means to quantitatively characterize the self-similarity of an object through relating the scaling of its segmented components to the level of magnification \cite{Mandelbrot}. The fractional or Fractal dimension indicates the degree of detail in this object and measures its complexity. It also demonstrates how much space it occupies between the Euclidean dimensions. The concept of Fractal dimension can be generalized to the non-discrete Fractals to define the "Fractal distribution" as

%----------------------------------------------------
\begin{equation}\label{Eq:Fractal-Dimention}
D_{f}(\alpha) = \frac{\partial log~f(\alpha)}{\partial log~(\alpha)}.
\end{equation}
%-------------------------------------------
The dimension of $D_{f}(\alpha)$ is a kind of a local dimension and those Fractals that respond to this local dimension,  are called "Multi-Fractals". Many Fractals in nature are not mathematically ideal Fractals and usually have Fractal structures only within
a certain region of magnification. Indeed, one often can find some regions where the system behaves like an ideal Fractal and can be described by some constant dimensions. The region of  " Mono-fractality" is what  we are interested in and we focus our analysis in this region, $D_{f}(\alpha)$ = D. Therefore, the Fractal distribution with fixed exponent (constant D) can be written as

%------------------------
\begin{equation}\label{Eq:density-function1}
log~f(\alpha) = D~.~log~\alpha+D_{0},
\end{equation}
%------------------------
where D$_{0}$ defines the normalization factor and $f(\alpha)$ is a power-law function and  $D$ parameter defines the Fractal dimension related to the Fractal distribution. For two independent magnification factors, $\alpha$ and $\beta$, this relation can be extended as
%------------------------
\begin{equation}\label{Eq:density-function2}
log~f(\alpha, \beta) = D_{\alpha}~.~log~\alpha+D_{\beta}~.~log~\beta+D_{\alpha\beta}~.~log~\beta~log~\alpha+D_{0},
\end{equation}
%------------------------
where $D_{\alpha\beta}$ demonstrates the dimensional correlation related to two magnification factors, $\alpha$ and $\beta$. These magnification factors, as mentioned in Refs.~\cite{Lastovicka:2002hw,Lastovicka:2004mq},  have some properties, namely, they should be positive, nonzero and dimensionless.\\
To describe the parton distribution functions at low $\textit{x}$, we start with unintegrated parton distribution functions and try to model them with a Fractal  distribution. The simplest unintegrated parton distribution functions or transverse  momentum distributions, TMDs, is the unpolarized distribution function $q(x,k_t)$. In a fast moving nucleon, it describes the probability of finding a quark carrying the longitudinal momentum fraction $\textit{x}$ of the nucleon's momentum and a transverse  momentum
$k_t= \left|  \overrightarrow{k_t} \right|$. It is related to the integrated parton distribution functions, PDFs, by
$q(x)=\int q(x,k_T) d^2 k_t$. For more detail see  \cite{Bastami:2020rxn}.\\
 It is worth noting that the distributions described by Eq. (10) have linear behavior in log-log space. It is shown that this linear behavior observed  for unintegrated sea quarks densities at low $\textit{x}$ and $k_t^2$. However, such linear behavior did not approved for gluon transverse momentum distribution as a function of  $k_t^2$ and it has only the self-similar behavior at low $\textit{x}$ region. For more detail see\cite{Jung:2006ji,Hansson:2007de,Knutsson:2008qs,Abdulov:2021ivr}. \\
Finally, because the Fractal dimensions are dimensionless, we choose $1+\frac{k_{t}^{2}}{q_{0}^{2}}$ instead of $k_{t}^{2}$ as a magnification factor related to these Fractal densities in $k_t^2$ space. In addition,  when the structure is probed deeper, \textit{x} goes to zero while the magnification factor should be increased. Therefore, we choose $\frac{1}{x}$ instead of \textit{x} in $\textit{x}$ space. Consequently, we choose $1+\frac{k_{t}^{2}}{q_{0}^{2}}$ and $\frac{1}{x}$ as magnification factors of the unintegrated parton densities in the low $\textit{x}$ and $k_t^2$ regions.\\
 To perform our analysis, we used the unintegrated parton distribution function of the sea quarks ${f}_{q/p} (x, k_{t}^2)$ and the unintegrated  gluon distribution ,$x{f}_{g/p}(x, k_{t}^2)$), as follows

\begin{eqnarray}\label{Eq:unintegrated-PDF}
log({f}_{q/p} (x, k_{t}^2))&=&D_{1}^{q} log(\frac{1}{x})log(1+\frac{{k_{t}^{2}}}{q_{0}^{2}})+D_{2}^{q} log(\frac{1}{x})+D_{3}^{q}log(1+\frac{{k_{t}^{2}}}{q_{0}^{2}})+D_{0}^{q}-log(M^{2}),  \nonumber  \\
x{f}_{g/p}(x, k_{t}^2)&=A &(\frac{1}{x})^{B^{g}}(1-x)^{C^{g}}(1-D^{g}x)~.~e^{{-\frac{(\mu-k_{t})^{2}}{\sigma^{2}}}} \frac{1}{M^{2}},
\end{eqnarray}
where, $k_{t}$ is the transverse momentum of the interacting partons. Here the sea quark distribution functions is kind of Fractal distributions with two magnification factors.  As stated earlier, due to the non linear log-log plot of the unintegrated gluon distribution, we used the Fractal approach to parameterize the gluon distributions only at low $\textit{x}$ region.  Eq. (12) shows that the gluon distribution function obeys the Fractal distribution at low $\textit{x}$ region. When $x \rightarrow 0$, the gluon density behaves as $x{f}_{g/p}(x, k_{t}^2) \sim x^{-B_g}$ which is a kind of Fractal distribution and has a linear behavior in log-log space. In this equation, The parameters $D_1^q$, $D_2^q$, and $D_3^q$ in the sea quark distribution have the following significance:
		$D_2^q$ controls the scaling behavior with 1/$x$, which characterizes the Fractal-like self-similarity as a function of magnification in $x$. A non-zero value of $D_2^q$ indicates the distribution scales as a power-law with 1/$x$ at low-$x$.
		$D_1^q$ provides an additional term accounting for the $x$ dependence that is independent of the Fractal scaling captured by $D_2^q$. This allows for deviations from perfect Fractal scaling.
		$D_3^q$ governs the scaling behavior with $k_t$, the other Fractal dimension related to spatial self-similarity as a function of transverse momentum magnification. A non-zero $D_3^q$ means the distribution scales as a power-law with $k_t$.
		Together, $D_2^q$ and $D_3^q$ encode the two-dimensional Fractal scaling properties, while $D_1^q$ allows for corrections. $D_2^q$ and $D_3^q$ would be expected to take on non-trivial values if the sea quark distribution truly has an underlying Fractal structure. The values of these parameters determined from fits to data can thus provide insights into the degree and nature of Fractality in the sea quark component of the proton.
So in summary, these parameters control the key scaling behaviors anticipated for a Fractal distribution, offering a way to quantitatively probe Fractal characteristics of the sea quarks.
The given parameterization for the gluon transverse momentum distribution has also been adopted in several other studies \cite{Hautmann:2014uua,Hautmann:2013tba,Lipatov:2022doa}.\\
Since the parton density functions of the sea quark and gluon  are dimensionless, therefore the two terms of $log M^{2}$ and $\frac{1}{M^{2}}$ are added to unintegrated sea quark and gluon densities, respectively \cite{Jahan:2014sqa}. We finally set $M^{2}=1~GeV^{2}$. We used the relation $xf_{i/p} (x, Q_0^2)=  \int_{0}^{Q_0^{2}}xf_{i/p} (x, k_{t}^2)~dk_{t}^2$ to obtain the  initial sea and gluon  PDFs.  Then,  we have

\begin{eqnarray}\label{eq:input}
x{f}_{q/p} (x, Q_{0}^2) & = & \frac{e^{D_{0}^{q}}q_{0}^{2}x^{-D_{2}^{q}+1}}{M^{2}(1+D_{3}^{q}-D_{1}^{q}log(x))}(x^{-D_{1}^{q}log(1+\frac{Q_{0}^{2}}{q_{0}^{2}})}(1+\frac{Q_{0}^{2}}{q_{0}^{2}})^{D_{3}^{q}+1}-1),  \nonumber  \\
x{f}_{g/p} (x, Q_{0}^2) & = & A'(\frac{1}{x})^{B^{g}}(1-x)^{C^{g}}(1-D^{g}x),
\end{eqnarray}
where $A'$ is:
\begin{equation}\label{Eq:A'}
A'= \frac{A}{M^2}(e^{-\frac{(\mu-1)^{2}}{\sigma^{2}}}+e^{-\frac{\mu^{2}}{\sigma^{2}}})\sqrt{\pi}\mu(erf(\frac{1-\mu}{\sigma})+erf(\frac{\mu}{\sigma})).
\end{equation}
 $erf(z)$ is the "error function" which is defined by \cite{error}
\begin{equation}\label{Eq:Erf}
erf(z)\equiv \frac{2}{\pi}\int_{0}^{z}e^{-t^{2}}dt.
\end{equation}
The 11 free parameters in initial PDFs are then determined by fitting them to the combined HERA experimental data. However, two of them are fixed and there are indeed only 9 fitted parameters.  How to do that is the subject of the next section.

%PLOTS 2
\begin{figure}[htb]
	\begin{center}
		\vspace{0.50cm}
		\resizebox{0.9\textwidth}{!}{\includegraphics{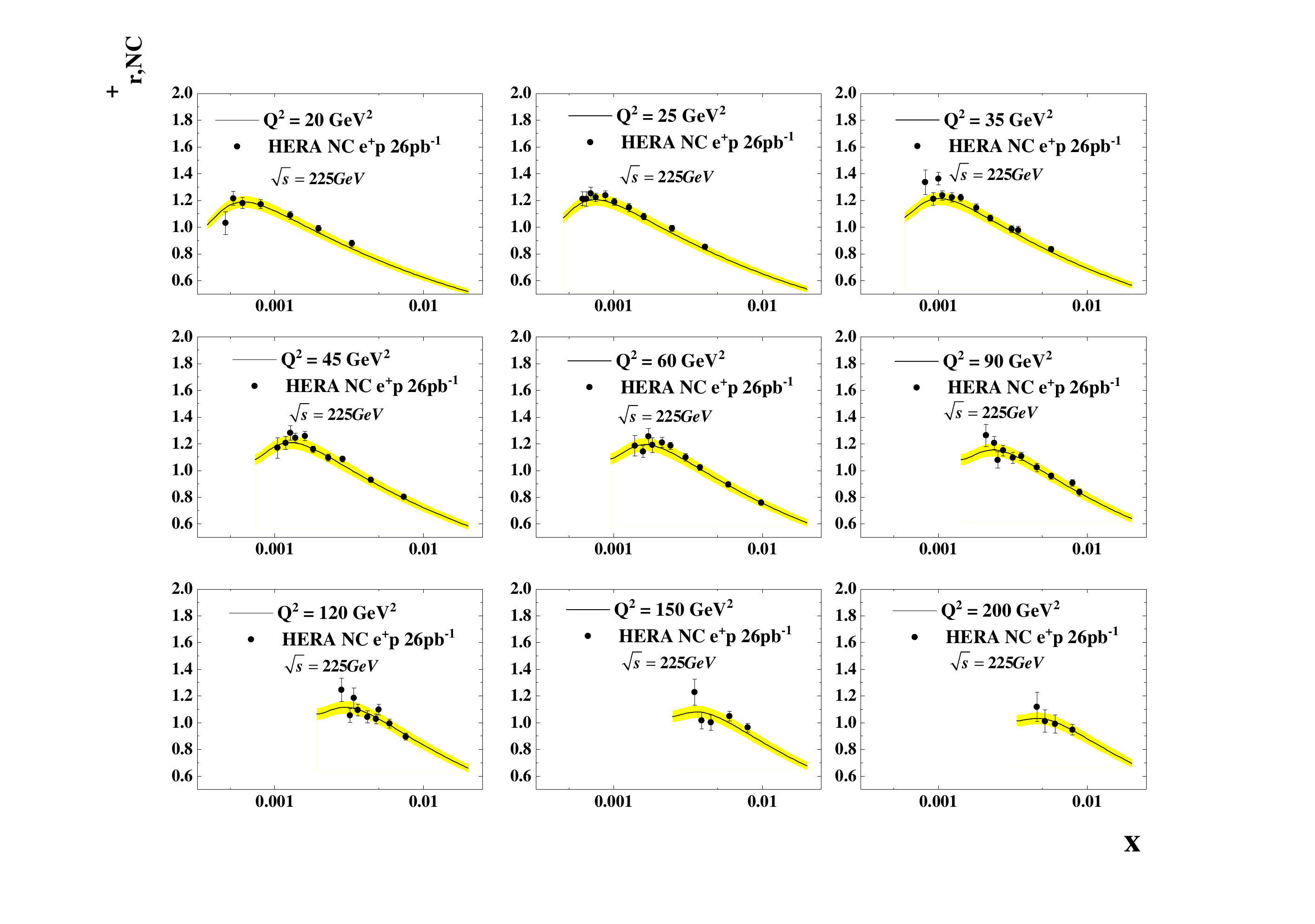}}
		\caption{{\small  The prediction for the reduced cross- section in the low x region below $x<0.01$ for $Q^2=20 GeV^2$ to
				$Q^2=200 GeV^2$  and  $\sqrt{s}=225GeV$. Data points are from NC interactions in HERA  positron- proton DIS processes.} \label{fig:fig6}}
	\end{center}
\end{figure}

\begin{figure}[htb]
	\begin{center}
		\vspace{0.50cm}
		\resizebox{0.9\textwidth}{!}{\includegraphics{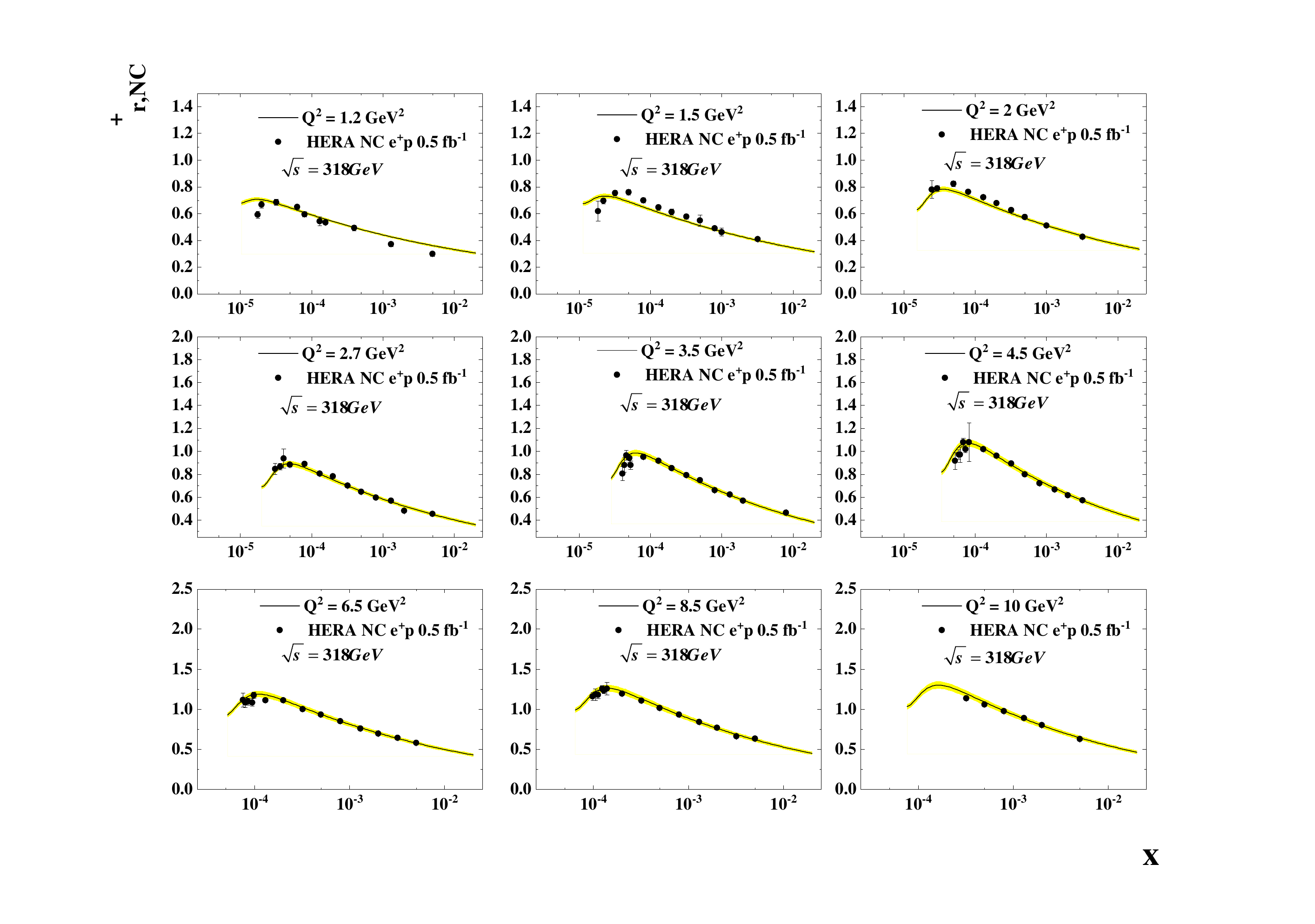}}
		\caption{{\small The prediction for the reduced cross- section in the low x region below $x<0.01$ for $Q^2=1.2 GeV^2$ to 	$Q^2=10 GeV^2$  and  $\sqrt{s}=318GeV$. Data points are from NC interactions in HERA  positron- proton DIS processes.} \label{fig:fig7}}
	\end{center}
\end{figure}

\begin{figure}[htb]
	\begin{center}
		\vspace{0.50cm}
		\resizebox{0.9\textwidth}{!}{\includegraphics{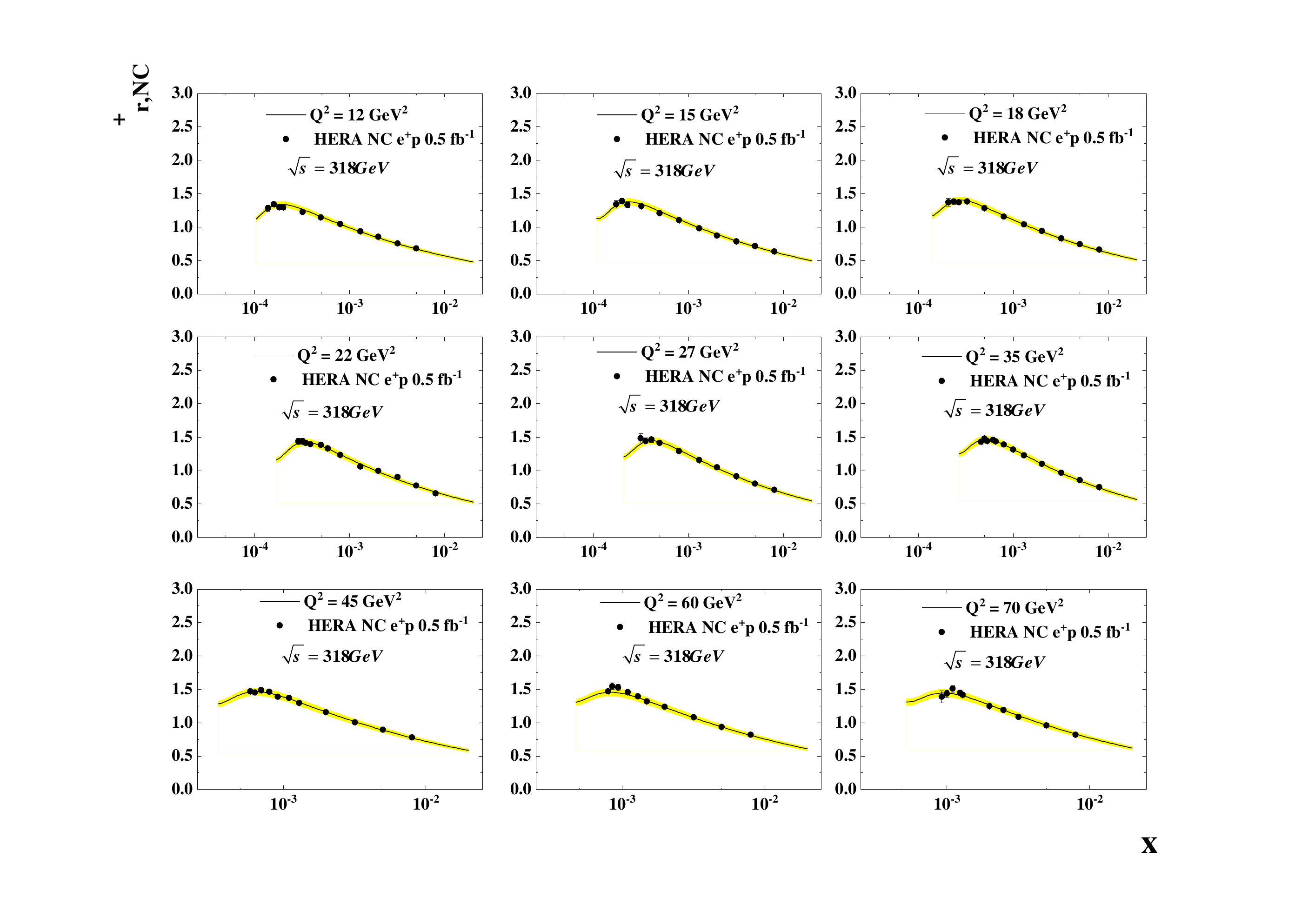}}
		\caption{{\small  The prediction for the reduced cross- section in the low x region below $x<0.01$ for $Q^2=12 GeV^2$ to 	$Q^2=70 GeV^2$  and  $\sqrt{s}=318GeV$. Data points are from NC interactions in HERA  positron- proton DIS processes.}  \label{fig:fig8}}
	\end{center}
\end{figure}

\begin{figure}[htb]
	\begin{center}
		\vspace{0.50cm}
		\resizebox{0.9\textwidth}{!}{\includegraphics{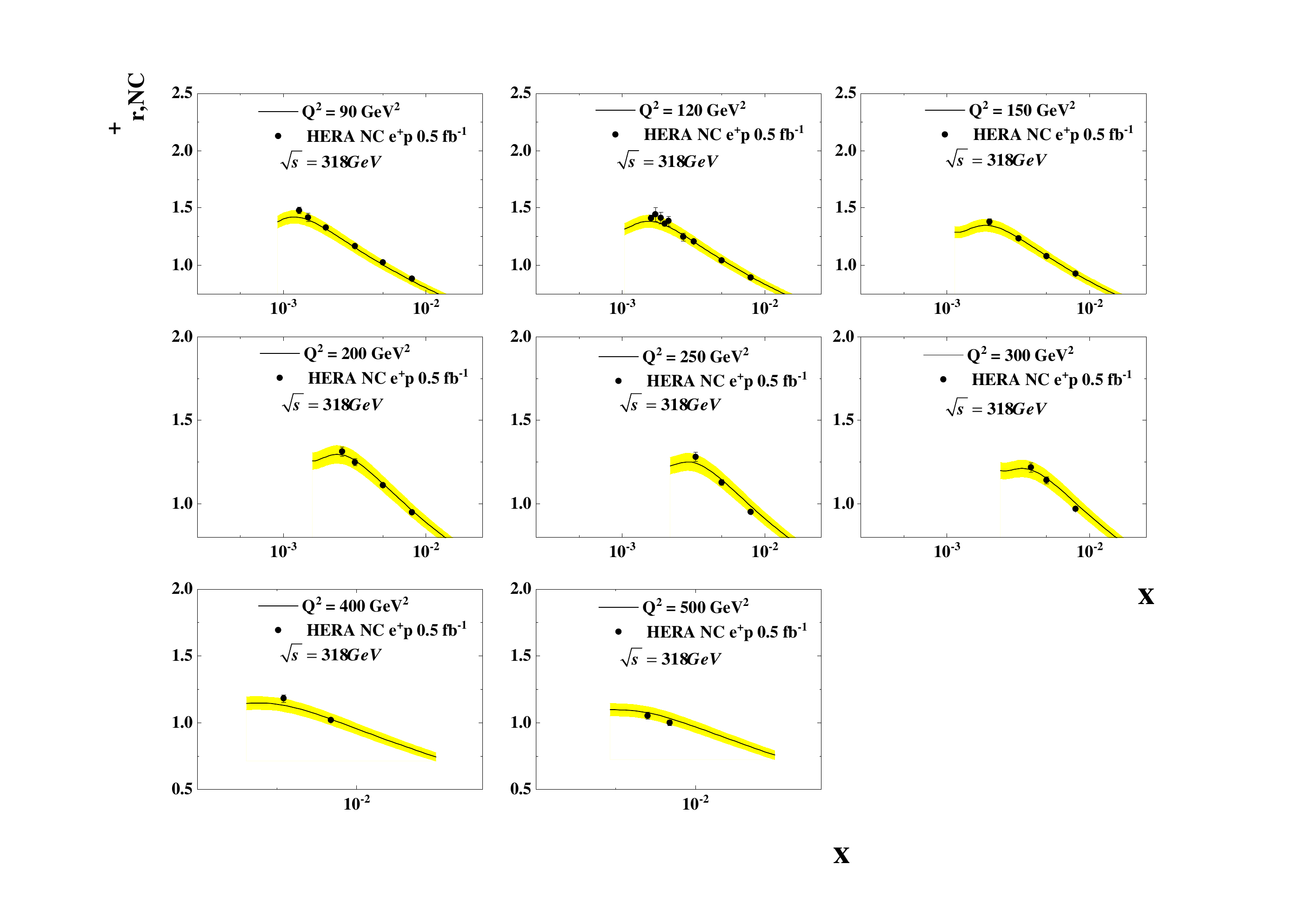}}
		\caption{{\small  The prediction for the reduced cross- section in the low x region below $x<0.01$ for $Q^2=90 GeV^2$ to 	$Q^2=500 GeV^2$  and  $\sqrt{s}=318GeV$. Data points are from NC interactions in HERA  positron- proton DIS processes.} \label{fig:fig9}}
	\end{center}
\end{figure}

\begin{figure}[htb]
	\begin{center}
		\vspace{0.50cm}
		\resizebox{0.9\textwidth}{!}{\includegraphics{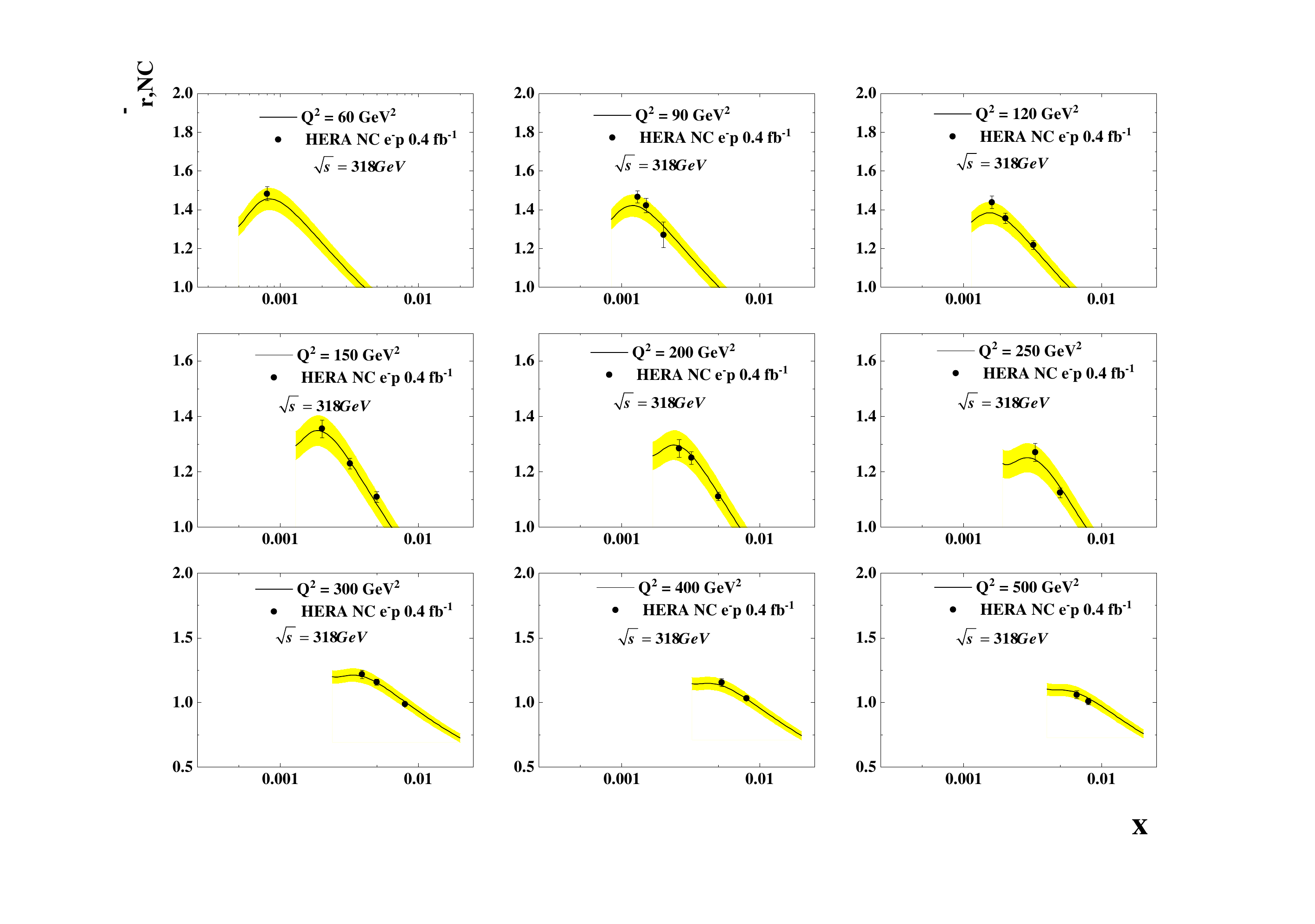}}
		\caption{{\small  The prediction for the reduced cross- section in the low x region below $x<0.01$ for $Q^2=60 GeV^2$ to 	$Q^2=500 GeV^2$  and  $\sqrt{s}=318GeV$. Data points are from NC interactions in HERA  electron- proton DIS processes.} \label{fig:fig10}}
	\end{center}
\end{figure}

\begin{figure}[htb]
	\begin{center}
		\vspace{0.50cm}
		\resizebox{0.9\textwidth}{!}{\includegraphics{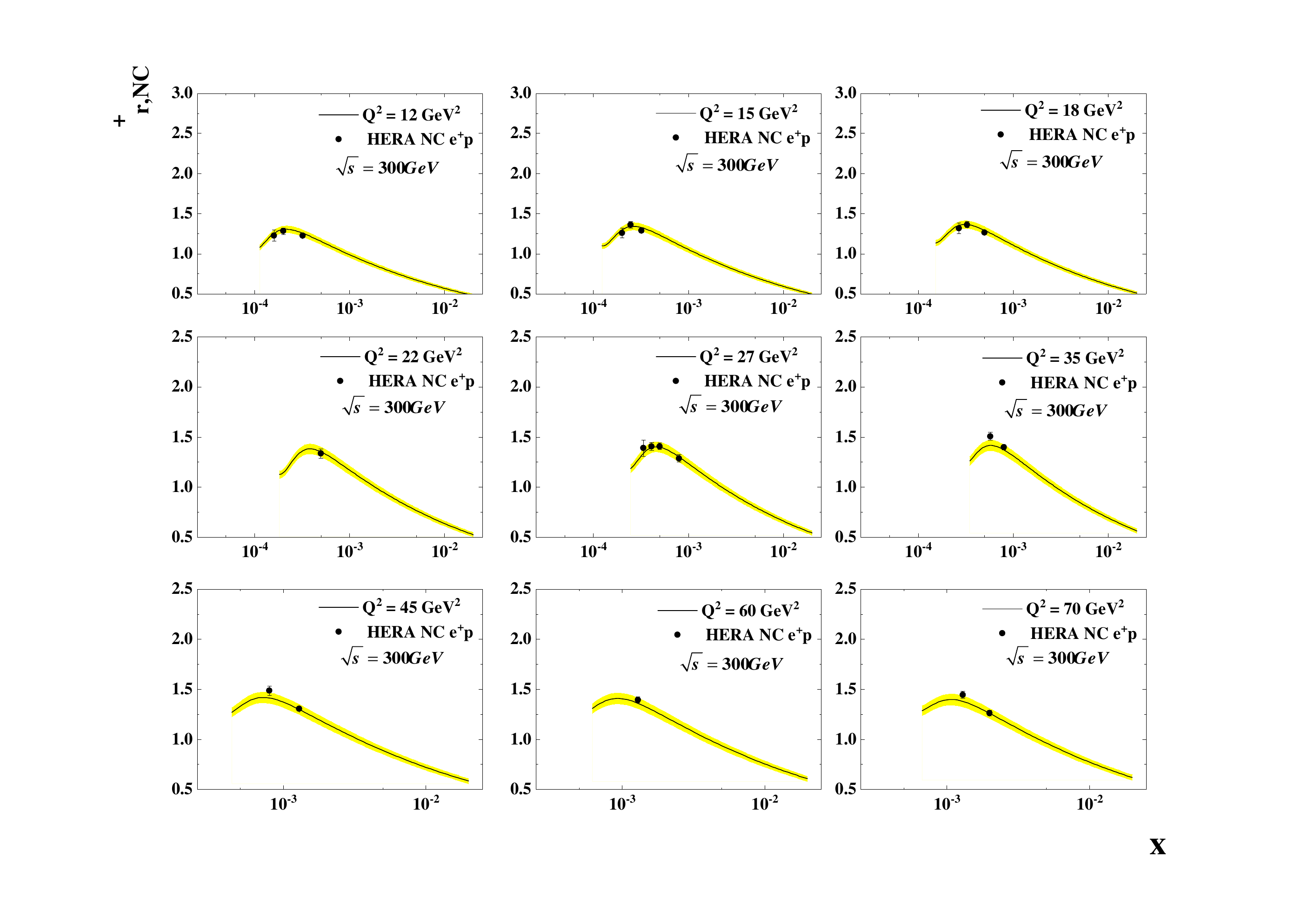}}
		\caption{{\small  The prediction for the reduced cross- section in the low x region below $x<0.01$ for $Q^2=12 GeV^2$ to 	$Q^2=70 GeV^2$  and  $\sqrt{s}=300GeV$. Data points are from NC interactions in HERA  positron- proton DIS processes.} \label{fig:fig11}}
	\end{center}
\end{figure}

\begin{figure}[htb]
	\begin{center}
		\vspace{0.50cm}
		\resizebox{1.0\textwidth}{!}{\includegraphics{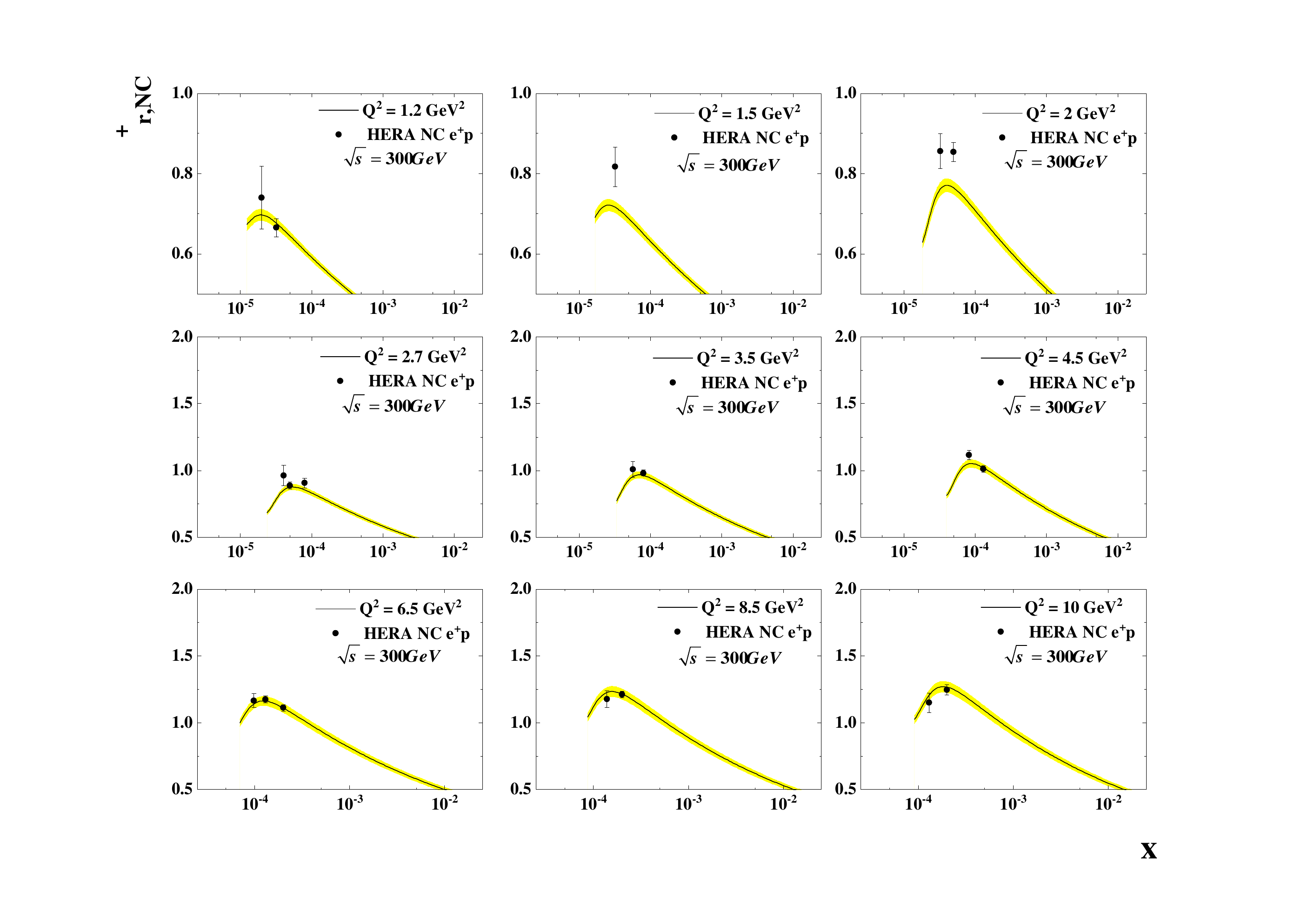}}
		\caption{{\small  The prediction for the reduced cross- section in the low x region below $x<0.01$ for $Q^2=1.2 GeV^2$ to 	$Q^2=10 GeV^2$  and  $\sqrt{s}=300GeV$. Data points are from NC interactions in HERA  positron- proton DIS processes.} \label{fig:fig12}}
	\end{center}
\end{figure}

% new plots

\begin{figure}[htb]
	\begin{center}
		\vspace{0.50cm}
		\resizebox{1.0\textwidth}{!}{\includegraphics{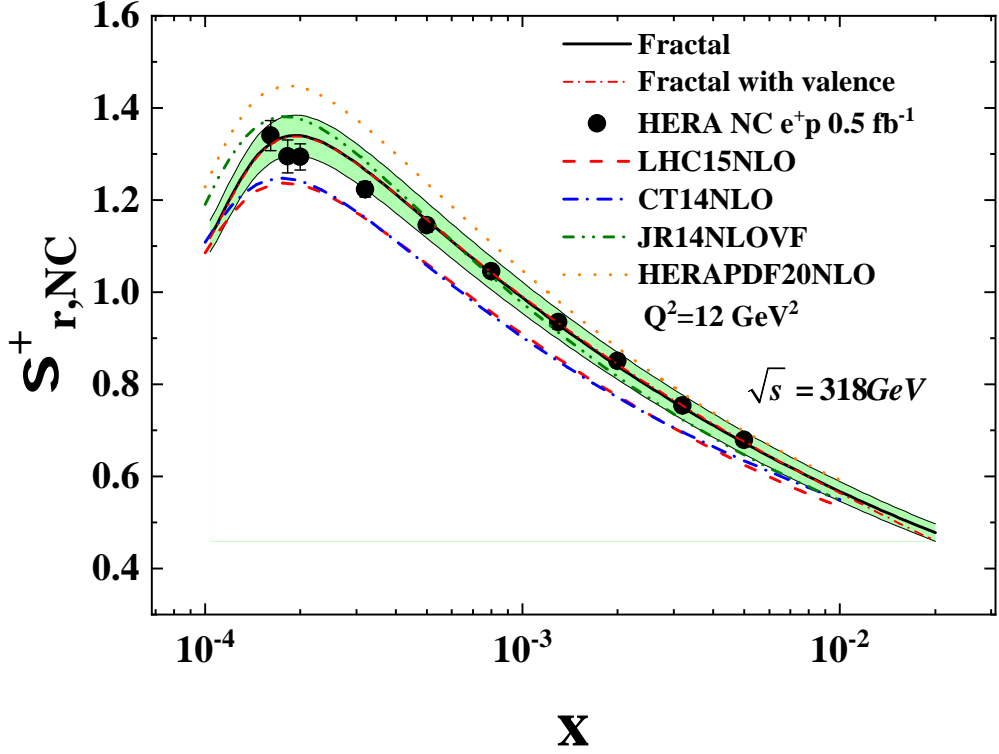}}
		\vspace{-10cm}
		\caption{{\small  The prediction for the reduced cross- section in the low x region below $x<0.01$ for $Q^2=12 GeV^2$ and  $\sqrt{s}=318GeV$ in comparison with experimental data \cite{H1:2015ubc,H1:2010fzx} and predictions from other phenomenological models \cite{Dulat:2015mca,Jimenez-Delgado:2014twa,Butterworth:2015oua,H1:2015ubc} to study the impact of the valence distributions at low x.} \label{fig:fig13}}
	\end{center}
\end{figure}

\begin{figure}[htb]
	\begin{center}
		\vspace{0.50cm}
		\resizebox{0.5\textwidth}{!}{\includegraphics{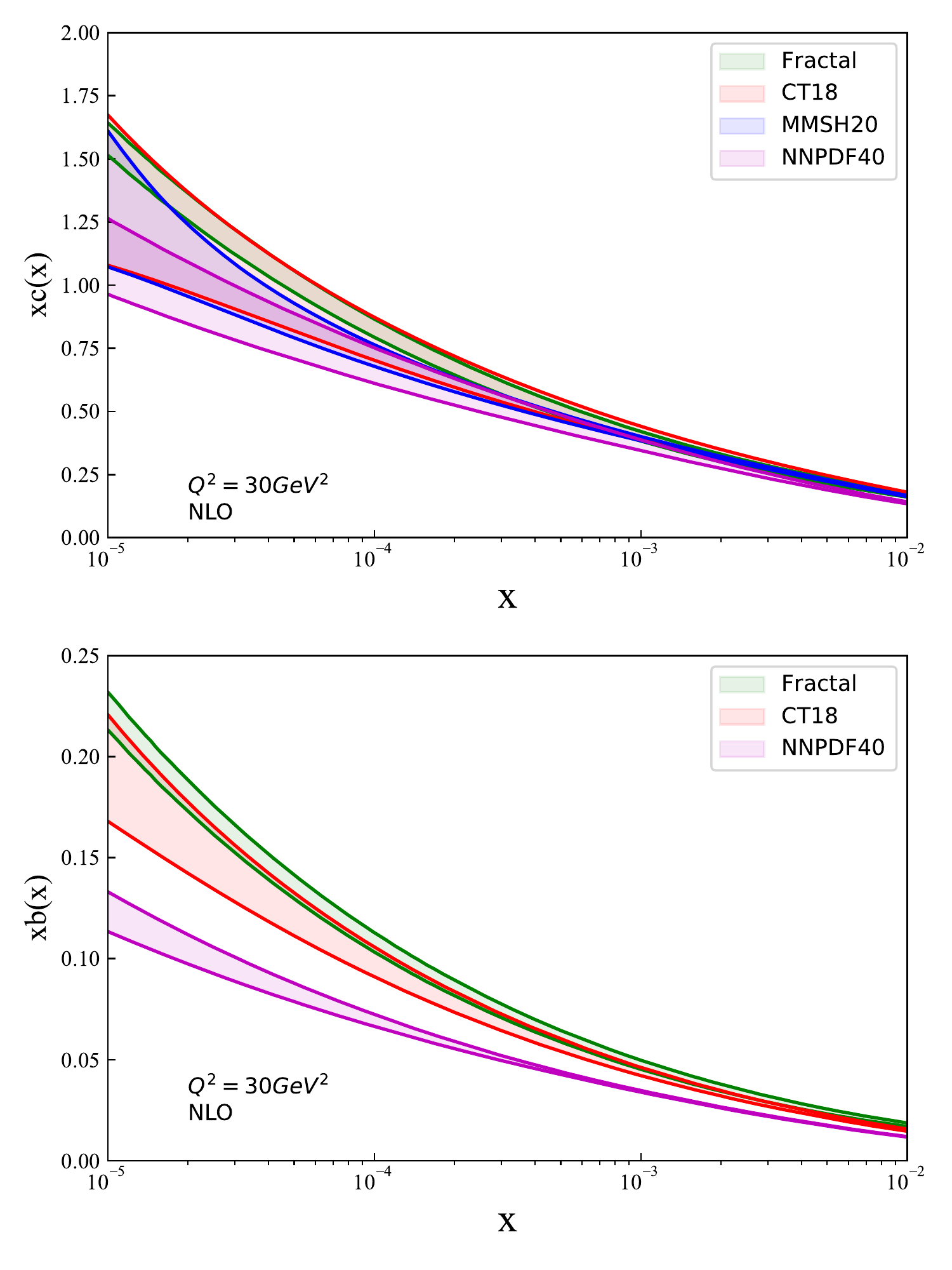}}
		\caption{{\small  The c and b quark distribution functions in comparison with those from CT18, MMSH20, and NNPDF4.0. parametrizations at $Q^2=30 GeV^2$ below to $x<0.01$ \cite{Bailey:2020ooq, NNPDF:2021njg,Hou:2019qau}. } \label{fig:fig14}}
	\end{center}
\end{figure}

\begin{figure}[htb]
	\begin{center}
		\vspace{0.50cm}
		\resizebox{1.0\textwidth}{!}{\includegraphics{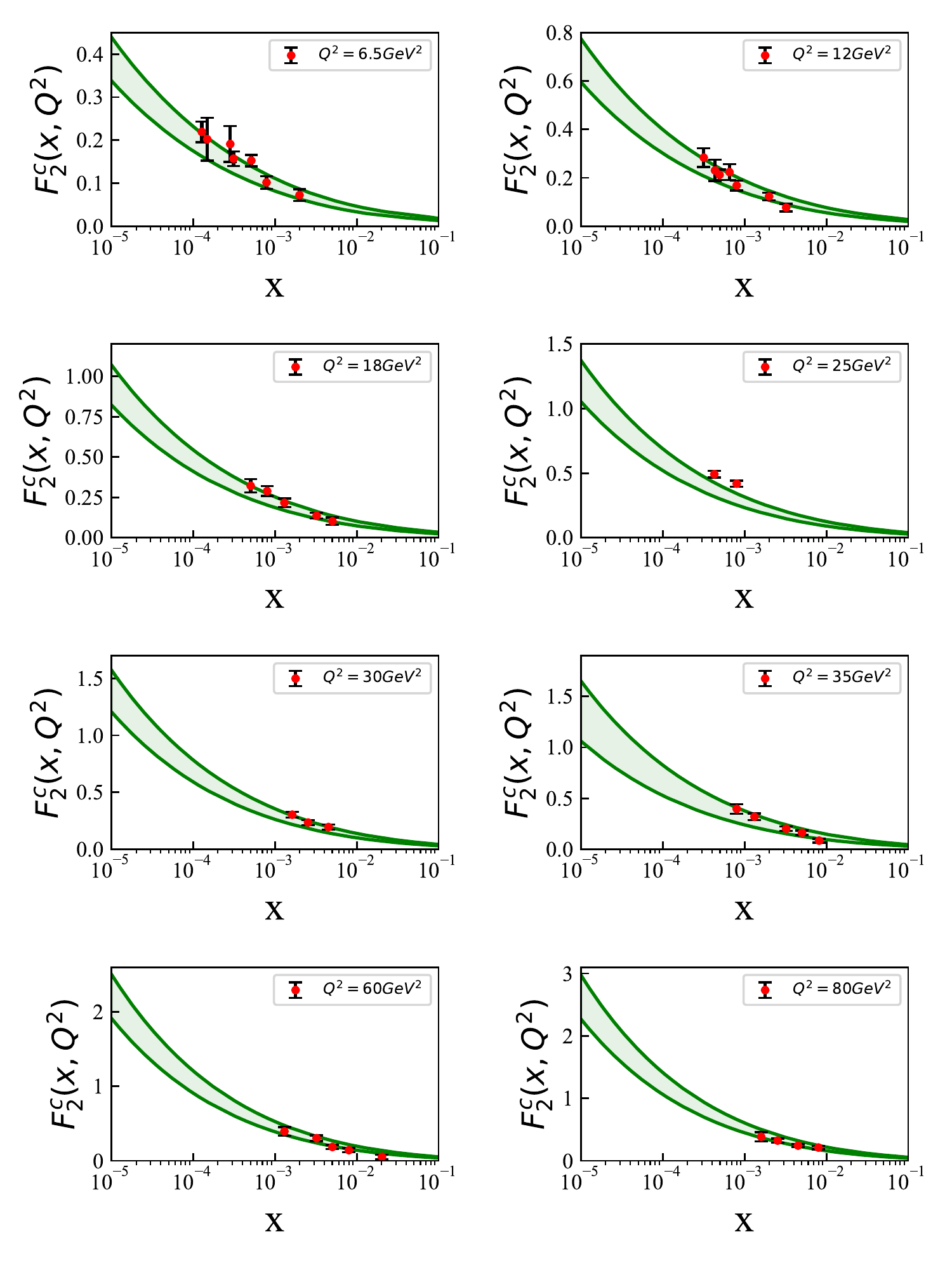}}
		\caption{{\small  The prediction for the c quark structure function $F_2^c(x,Q^2)$  in the low x region for various values of $Q^2$  and comparison with experimental data from HERA \cite{H1:2011unn,H1:2009uwa}.} \label{fig:fig15}}
	\end{center}
\end{figure}

\begin{figure}[htb]
	\begin{center}
		\vspace{0.50cm}
		\resizebox{1.0\textwidth}{!}{\includegraphics{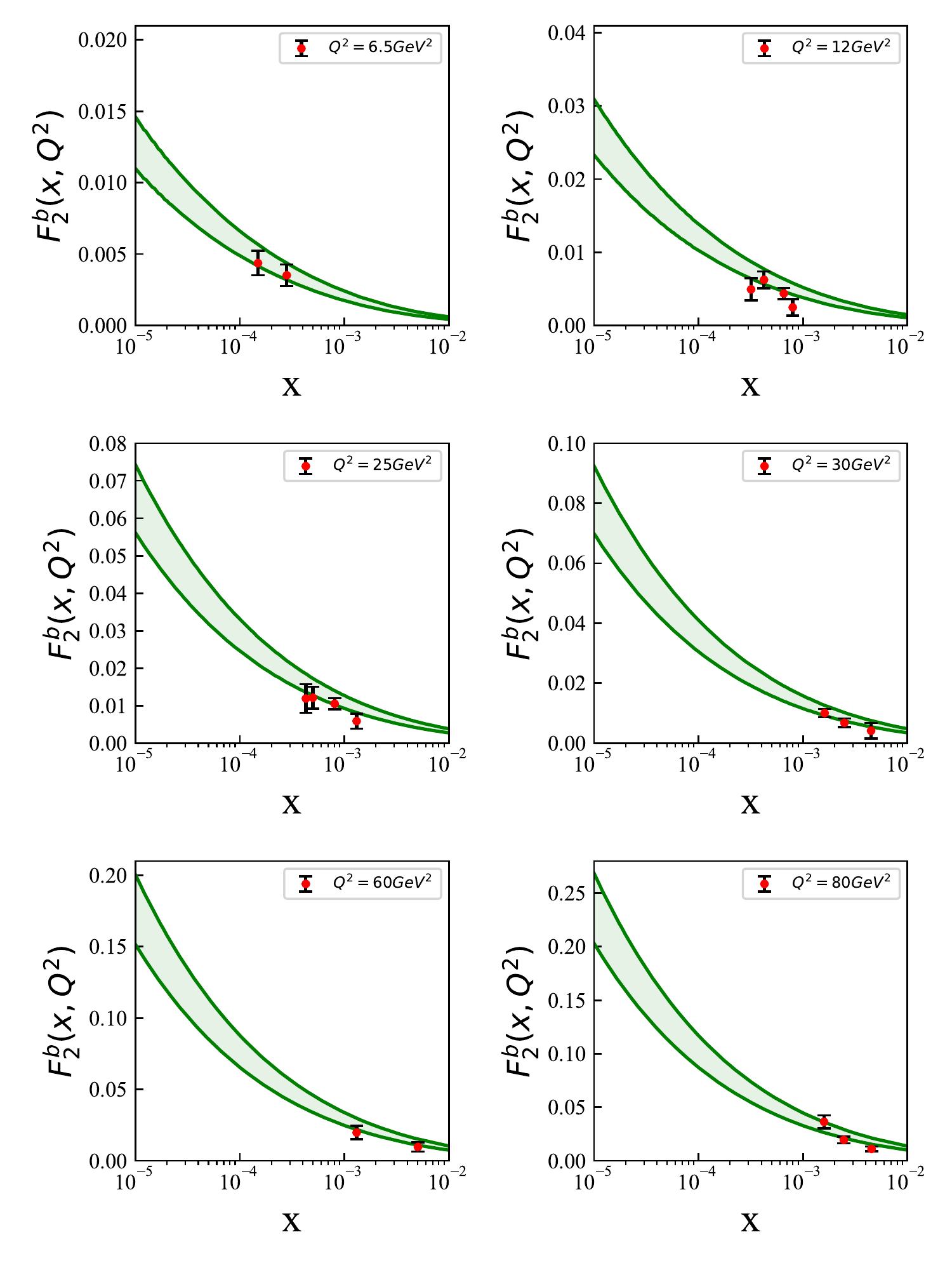}}
		\caption{{\small  The prediction for the b quark structure function $F_2^b(x,Q^2)$  in the low x region for various values of $Q^2$  and comparison with experimental data from HERA \cite{H1:2011unn,H1:2009uwa}.} \label{fig:fig16}}
	\end{center}
\end{figure}

% FL PLOTS

\begin{figure}[htb]
	\begin{center}
		\vspace{0.50cm}
		\resizebox{1.0\textwidth}{!}{\includegraphics{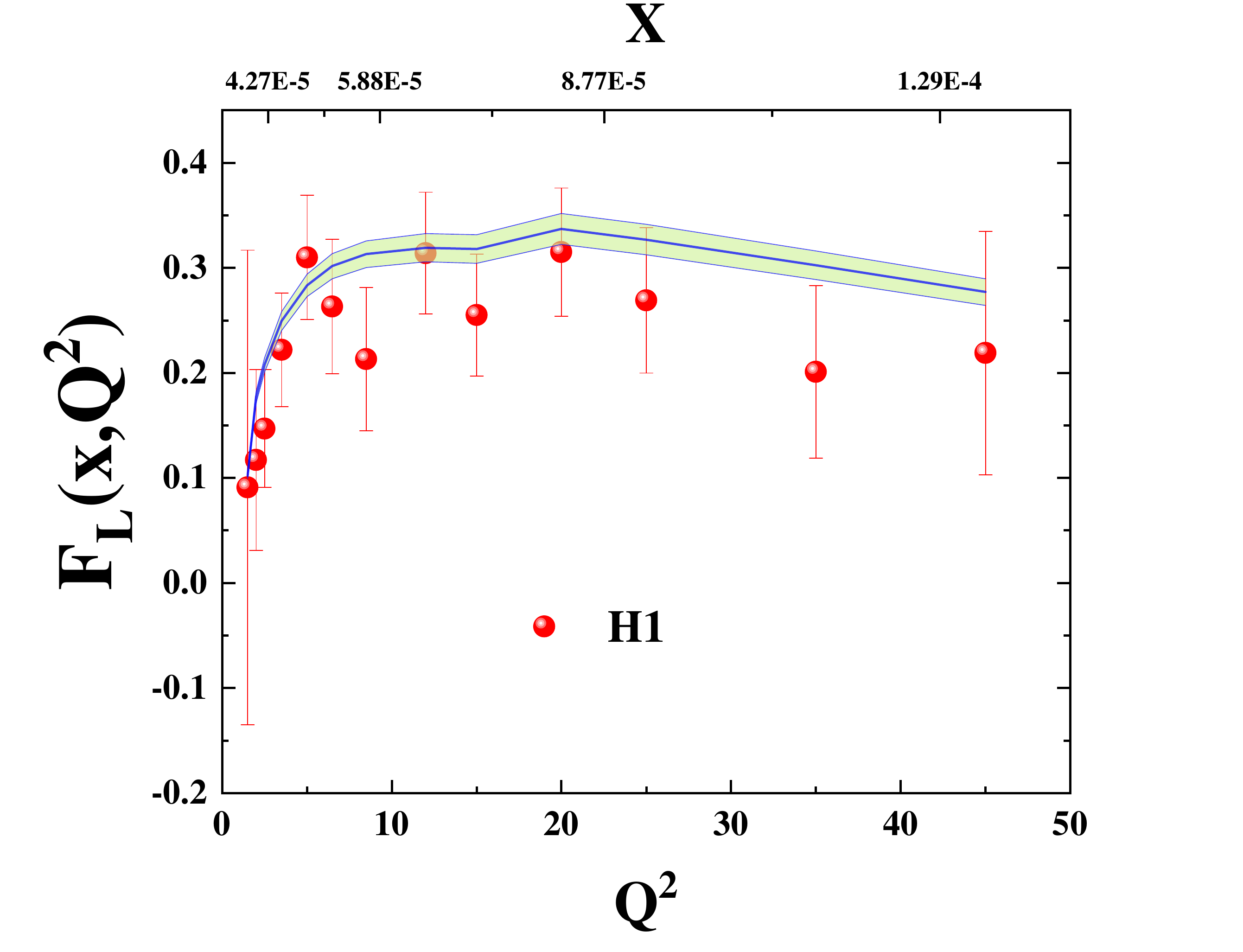}}
		\caption{{\small  The $F_L$  structure function  in the low x region for various values of $Q^2$  and comparison with experimental data from HERA. This data is used in the analysis} \label{fig:fig17}}
	\end{center}
\end{figure}

\begin{figure}[htb]
	\begin{center}
		\vspace{0.50cm}
		\resizebox{1.0\textwidth}{!}{\includegraphics{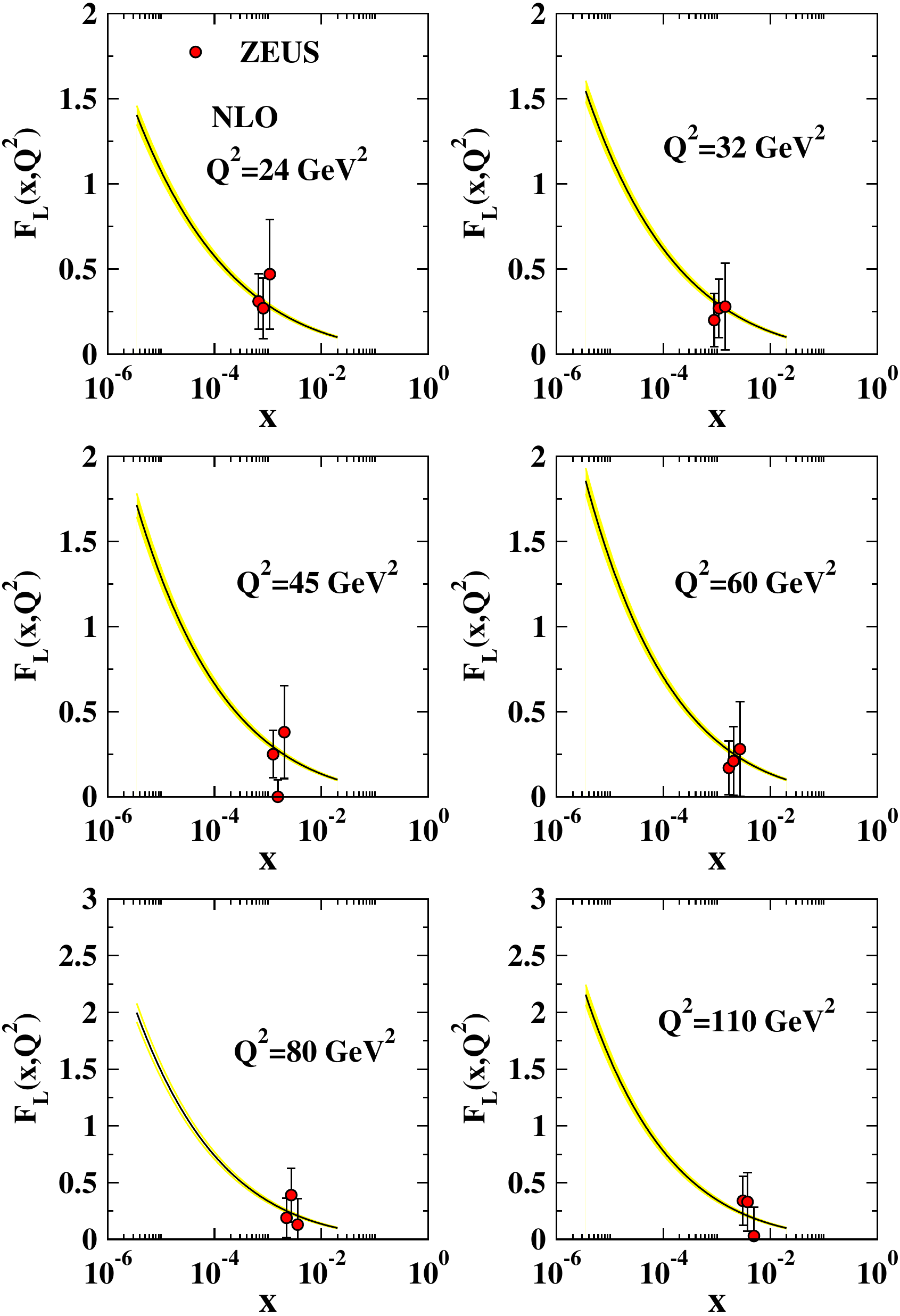}}
		\caption{{\small   The prediction for  $F_L$  structure function  in the low x region for various values of $Q^2$  and comparison with experimental data from HERA. This data is "not" used in the analysis.} \label{fig:Graph18}}
	\end{center}
\end{figure}

%
%%%%%%%%%%%%%%%%%%%%%%%%%%%%%%%%%%%%%%%%%%%%%%%%%%%%%%%%%%%%%%%%%%%%%%%
\subsection{ Experimental observable and the minimization processes } \label{sec:data-selection}
%%%%%%%%%%%%%%%%%%%%%%%%%%%%%%%%%%%%%%%%%%%%%%%%%%%%%%%%%%%%%%%%%%%%%%%
%

The experimental data presented in this section will be used in our analysis. In the next step, the error estimation process will be described.
 To study electron(positron)- proton scattering in DIS processes, we used HERA experimental data. In this analysis, we choose the subset of the combined HERA  data related to the NC interactions in low x region below to $x<0.01$ region \cite{H1:2015ubc,H1:2010fzx}. 
The total number of used experimental data is then $N_{data}=648$. They are summarized in Table. (I).\\
To determine the best values of unknown parameters related to the initial parton densities  in Eq.~\ref{eq:input}, one needs to minimize the $\chi^2$ with respect to free parameters in  initial PDFs. In QCD analysis of PDFs, the global goodness-of-fit procedure  follows the usual approach which defines $\chi^{2}_{\textrm{global}}(\{\xi_{i}\})$ as

%--------------------------------
\begin{equation}
\chi_{\textrm{global}}^2(\{\xi_{i}\})
= \sum_{m=1}^{m^{\textrm{exp}}} \,
w_{m}  \,
\chi_m^2 (\{\xi_{i}\})\,,
\label{eq:chi2-1}
\end{equation}
%--------------------------------
where $m$ run over various experiments. $w_{m}$ allows us,
to include datasets with different weight factors.
Each experiment
contributes with $\chi_m^2(\{\xi_{i}\})$ to the global $\chi^2$.
These terms depend on the fit parameters $(\{\xi_{i}\})$, of the proton PDFs at
the initial scale. $\chi_m^2(\{\xi_{i}\})$ is  calculated as follows
%--------------------------------
\begin{eqnarray}
\chi_m^2(\{\xi_{i}\}) &=&
\left(\frac{1-{\cal N}_m}
{\Delta{\cal N}_m}\right)^2 +
\nonumber \\
\hspace{-0.5cm}\sum_{j=1}^
{N_m^{\textrm{data}}} &&
\left(\frac{({\cal N}_m  \,
{\cal O}_j^{\textrm{data}}-
{\cal T}_j^{\textrm{theory}}
(\{\xi_i\})}{{\cal N}_m \,
\Delta_j^{\textrm{data}}}\right)^2 \,,
\label{eq:chi2-2}
\end{eqnarray}
%----------------------------------
where $j$ runs over data points, $m$ indicates each
datasets, and $N^{\textrm{data}}_m$ corresponds to the
total number of data points used in the analysis. Here we work with only one data set from HERA, then $N_m=1$. In the above equation,
${\cal O}^{\textrm{data}}_{j}$ refers to the value of the measured data
point for a given observable, and $\Delta^{\textrm{data}}_j$
is the total experimental error calculated from the statistical and
systematic errors added in quadrature. The theoretical predictions
for each data point $j$ are represented by
${\cal T}^{\textrm{theory}}_{j}(\{\xi_{i}\})$, which has to be
calculated at the same experimental kinematic point $\textit{x}$ and
$Q^{2}$ using the DGLAP evolution equations for the PDFs with given parameters
$(\{\xi_{i}\})$.
%--------------------------------
%------------------
To obtain the best parametrization of the  PDFs, we used
the widely-used CERN program library {\tt MINUIT}~\cite{James:1994vla}. An important task in a QCD analysis of PDFs is the estimation of uncertainties of the PDFs obtained from the $\chi^2$ optimization.
There are  well-defined procedures for propagating experimental uncertainties on the fitted data points through to the PDFs uncertainties. Here we used the  ``Hessian method'' \cite{Pumplin:2001ct}, for estimating uncertainties of the PDFs at low x region.
The ``Hessian method'' was widely used in many global QCD analyse of PDFs. It is based on linear error propagation and involves the production of eigenvector PDFs sets suitable for convenient use by the end user.  In addition the  Hessian approach, which is based on the covariance matrix diagonalization, provides us a simple and efficient method for calculating the uncertainties of PDFs at low x. Following that, an error estimation analysis can be obtained by using the ``Hessian method''  which is determined by the CERN program library MINUIT. For more detail about the Hessian method see \cite{Pumplin:2001ct}.
Finally, the uncertainties of PDFs as well as the related observable are estimated using the ``Hessian matrix'' and their values at higher $Q^2$ ($Q^2  > Q_0^2$) are calculated using the DGLAP evolution equations. The QCD evolution of the PDFs are given by the QCDNUM program \cite{Botje:2010ay}.

%%%%%%%%%%%%%%%%%%%%%%%%%%%%%%%%%%%%%%%%%%%%%%%%%%%%%%%%
%%%%%%%%%%%%%%%%%%%%%%%%%%%%%%%%%%%%%%%%%%%%%%%%%%%%%%%%
%%%%%%%%%%%%%%%%%%%%%%%%%%%%%%%%%%%%%%%%%%%%%%%%%%%%%%%%

%----------------------------------
\begin{table*}[htb]
\caption{List of all DIS data points above $Q^2$ = 1.0 GeV$^2$ used in this  analysis \cite{H1:2015ubc,H1:2010fzx}.} \label{tab:tabdata}
\begin{tabular}{l c c c c c c}
\hline
DIS experiment (NC) &  [$x_{\rm min}, x_{\rm max}$]  & [Q$^2_{\rm min}$, Q$^2_{\rm max}$] GeV$^2$  & Number of data points
\tabularnewline
\hline\hline
$e^{-}p \rightarrow e^{-}X $  &  [$8 \times 10^{-4}$ -- $8.5 \times 10^{-3}$] &[60--650] & 27  \\ \hline		
$e^{+}p \rightarrow e^{+}X $  &   [$3.48 \times 10^{-5}$ -- $9.3 \times 10^{-3}$] & [1.5--400] & 135  \\	\hline
$e^{+}p \rightarrow e^{+}X $ &  [$2.79 \times 10^{-5}$ -- $9.3 \times 10^{-3}$] & [1.5--500] & 182   \\		\hline
$e^{+}p \rightarrow e^{+}X $ &   [$2 \times 10^{-5}$ -- $8.5 \times 10^{-3}$] & [1.2--650] & 52 	\\ \hline
$e^{+}p \rightarrow e^{+}X $ &  [$1.79 \times 10^{-5}$ -- $8.5 \times 10^{-3}$] & [1.2--650] & 239   \\ \hline
F$_{L}$  &  [$2.79 \times 10^{-5}$ -- $1.46 \times 10^{-3}$] &[1.5--45] & 13 \\	
\hline \hline
\multicolumn{1}{c}{\textbf{Total data}}  &   & & \textbf{648} &    \\  \hline
\end{tabular}	
\end{table*}

%
%-------------------

%----------------------------------
\begin{table}%[htb]
\begin{center}
\caption{ The best values, obtained for 9 free parameters in Eq.~(\ref{eq:input}) at the initial scale of Q$_0^2$ = 1 GeV$^2$. Values marked with $(*)$ are fixed in this analysis.} \label{tab:fit-parameters}
\begin{tabular}{|l|l|l|c|ccccccc}
	\hline
	Parameters & $ x{f}_{q(g)/p} (x, Q_{0}^2)$ & $ x{f}_{q(g)/p} (x, Q_{0}^2)$ with Valence contribution  \\
	   \hline
	$D_{0}^{q}$  & $4.81 \pm 0.01$ &    $   6.768 \pm  0.0102$   \\ \hline
	$q_{0}$  & $0.0187 \pm 0.0001$ &   $   0.024 \pm  0.0001$    \\ \hline
	$D_{1}^{q}$  & $-0.0051 \pm 0.0009$  &    $0.139 \pm  0.001$   \\ \hline
	$D_{2}^{q}$  & $1.138 \pm 0.002$   &    $0.937 \pm 0.002$   \\ \hline
	$D_{3}^{q}$  & $-1.285 \pm 0.005$    &    $-2.996 \pm 0.014$   \\ \hline
		$A$  & $0.191 \pm 0.005 $   &   $0.239 \pm 0.0058$   \\ \hline
	$B^{g}$  & $-3.01 \pm 0.03$   &     $-1.952 \pm 0.026$   \\ \hline
	$C^{g}$  & $4^{*}$   &     $4^{*}$   \\ \hline
	$D^{g}$  & $-3071.1 \pm 80.6$   &    $-908.36 \pm 22.04$   \\ \hline
	$\sigma$  & $7.15 \pm 6.18 $    &    $4.81 \pm 0.01$   \\ \hline
	$\mu$  & $1.5^{*} $     &     $1.5^{*} $   \\ \hline
	\hline
	$\frac{\chi^2}{d.o.f}$ & $1.148 $ & $1.198$   \\ \hline
	
	\hline
\end{tabular}
\\

\end{center}
\end{table}
%----------------------------------

%---------------------------------------------------------
%%%%%%%%%%%%%%%%%%%%%%%%%%%%%%%%%%%%%%%%%%%%%%%%%%%%%%%%%%%%%%%%%%%%%%%%%%%%%%%%%%%%%%%%%%%%%%%%%%%%%%%%%%%%%%%%%%%%%%%%%%%%%%%%%%%%%%%%%%%%%%%%%%%%%%%%%%%%%%%%%%%%%%%%%%%%%%%%%%%%%%%%%%%%%%%%%%%

\begin{table*}[htbp]
\caption{ The correlation matrix elements for the 9 free parameters at NLO fit \label{correlation-matrix-elements}}
\begin{tabular}{lcccccccccccc}
\hline
& & $q_{0}$  & $D_{0}^{q}$  &   $ B^{g}$  & $D_{1}^{q}$& $D_{2}^{q}$  &   $ D_{3}^{q}$   &   $\sigma $  & $A$ & $D^{g}$ \\     \hline
$q_{0}$ & & 1 &  &  & & & & & &   \\
$D_{0}^{q}$ & & -0.303 & 1 &  &      \\
$B^{g}$  && -0.051 &-0.052  & 1 &  & & & & &     \\
$D_{1}^{q}$ & & -0.144 & -0.142 & 0.042&  1  & & & & &   \\
$D_{2}^{q}$ & & -0.127 & -0.125 &  0.044& -0.726 & 1& & & &   \\
$D_{3}^{q}$ & & -0.312 &-0.314  &  -0.056& -0.123  &-0.103 &1 & & &   \\
$\sigma$ & & -0.039 & -0.040 & -0.293 & 0.040 &0.040 &-0.043 &1 & &   \\
$A$ & & -0.043 & -0.043 & -0.318 & 0.043&0.044 &-0.047 &-0.301 &1 &  \\
$D^{g}$ & & 0.043 & 0.043 &0.317 &-0.043&-0.044 &0.047 &0.300 &0.329 &1   \\  \hline
\end{tabular}
\end{table*}

%%%%%%%%%%%%%%%%%%%%%%%%%%%%%%%%%%%%%%%%%%%%%%%%%%%%%%%%%%%%%%%%%%%%%%%
\section{Results and  Discussion   } \label{sec:Results}
%%%%%%%%%%%%%%%%%%%%%%%%%%%%%%%%%%%%%%%%%%%%%%%%%%%%%%%%%%%%%%%%%%%%%%%
%

In the last sections, we discussed the possibility of studying the parton distributions in the low x region below $x<0.01$ using the Fractal distributions. In this section, we will present and discuss obtained results based  on the Fractal behavior of the TMDs with  analysis of the Neutral current HERA data on inclusive DIS processes at the low $x$ region.\\
With two scenarios, we take the Fractal approach. In the first one, we ignore the role of valence quark distributions at the low x below $x<0.01$ and work just with symmetric sea quarks and gluons in VFNS. In the second scenario,  we include the valence quark distribution  from MSHT20 PDFs \cite{Bailey:2020ooq}, and then rerun the program to study the impact of the valence quarks on the obtained results at the low x region.\\
 The best values, obtained for  9 free parameters in Eq.~(\ref{eq:input}) at the initial scale of Q$_0^2$ = 1 GeV$^2$ for two scenarios and the correlation matrix elements for the these parameters  for the first one are summarized in Tables. (II) and (III) respectively. In  Table (II), the values marked with $(*)$ are fixed in this analysis. The parametrization for the initial input PDFs  was presented in Sec. III in Eq.~(\ref{eq:input}).\\
The extracted  PDFs at low $\textit{x}$  below $x<0.01$ are plotted in  Fig.~(\ref{fig:inputpdfs}) at  $Q_0^2=1 GeV^2$ and   $Q^2=10 GeV^2$. At the initial scale of $Q_0^2$, the vertical line shows the validity range of the Fractal approach. The Fractal distributions describe the low x behavior of the PDFs below $x<0.01$.  The uncertainty bands for NLO QCD analysis at  $90 \% $ C.L. limit which are obtained using the same approach for the input parametrization and error propagation are also shown in this figure.  In  Fig.~(\ref{fig:pdfsq10}),  our results based on the Fractal approach are compared to those obtained by CT14, JR14 and LHC15  global analyses \cite{Dulat:2015mca,Jimenez-Delgado:2014twa,Butterworth:2015oua}. The results show a nice agreement between them at the low $\textit{x}$ region. It should be noted that the Fractal PDFs are valid only for $x<0.01$ and this region is determined by a vertical line in Fig.~(\ref{fig:inputpdfs}). In Fig.~(\ref{fig:pdfsq10ct}), we have another comparison of our results with those from CT14 PDFs, but now the uncertainties  added to the figure. This comparison shows that using the Fractal based PDFs  at low $\textit{x}$  considerably reduces the uncertainties.
The predictions for the reduced cross-section as a function of x in intervals of $Q^2$ between $1.2 GeV^2$ to $500 GeV^2$ and $\sqrt{s}$ between $225 GeV$ to $318 GeV$ are presented in figures 6-12. The results show a good agreement between experimental data \cite{H1:2015ubc,H1:2010fzx} and those predicted by the Fractal approach. This agreement is also seen at low $Q^2$ for $Q^2< 8.5 GeV^2$ and  between $x\cong 10^{-5}$ to  $x\cong 10^{-3}$. Note that the low $Q^2$ experimental data is almost excluded in many global analyses of PDFs.\\
Plots show that our results for the reduced cross-section in positron-proton DIS processes describe the data well. It is worth pointing out in the context of the Fractal approach, the results presented here clearly show that expected statistical accuracy are very good for all experimental data. This suggests that the Fractal approach for parametrizing the TMDs and PDFs at low x  may describes the low x (and low $Q^2$) data well. We believe that this agreement with experimental data is related to the initial inputs which are based on choosing the Fractal  distributions for the TMDs.
Originally, the self similarity exists for the transverse momentum distributions inside proton. The transverse momentum distributions for the sea quarks have self-similar or Fractal behavior   at low $\textit{x}$ and $k_t^2$   whereas the gluon transverse momentum distribution has this behavior only for low $\textit{x}$ region.\\
In Fig. 13 we plot the reduced cross-section in the low x region in comparison with those from experimental data and  HERAPDFs, LHC15,CT14 and JR14  parameterizations \cite{Dulat:2015mca,Jimenez-Delgado:2014twa,Butterworth:2015oua,H1:2015ubc}. This figure implies that for $x>0.01$, the reduced-cross section prediction is affected by including the valence quark distributions in the analysis. This is an interesting result and is compatible with the new study based on the consideration of the higher-twist effects in deep inelastic
scattering at HERA at moderate $Q^2$. This study is  shown that the higher-twist correction  affect significantly the fitted gluon and sea density functions at
low x region below $x < 0.01$ and moderate  $Q^2$, while
valence quark distributions are not sensitive to higher-twist effects and can be ignored in this region \cite{Motyka:2017xgk}.\\
In Fig. (14) we extracted the heavy quark distribution functions at the low x region and compared them with those from CT18 and NNPDF4.0 and MMSH20 parameterizations \cite{Bailey:2020ooq, NNPDF:2021njg,Hou:2019qau}.\\
In Fig. (15) and Fig. (16), we calculate the heavy quark structure functions $F_2^{c}(x, Q^2)$ and $F_2^{b}(x, Q^2)$ and compare them with experimental data from HERA \cite{H1:2011unn,H1:2009uwa} to check if the Fractal model can predict these structure functions appropriately. Note that the experimental data related to the heavy quarks are not included in the analysis.  According to the results, they are in good agreement. Finally, in Fig. (17) and Fig. (18), we compared the $F_L$ structure functions with experimental data used in the analysis and those ZEUS Experimental data that we didn't use in the study. The results clearly describe the low x region using the Fractal PDFs. In the end, the following issues could be addressed in  future works:
\begin{itemize}
	\item The experimental data related to the heavy quarks can included in the analysis to determine the free parameters of the model.
	\item The CCFM evolution equations \cite{Hautmann:2014uua} can be used to calculated the TMDs at the low x region as a function of $k_t^2$ for different values of $x$.
	\item  There are some experimental data related to the W-boson jet production at the LHC that sensitive to the TMDs at  low x region \cite{ATLAS:2012ikx}. The Fractal TMDs introduced here can be used to  compute predictions for the W-boson + jets final states.   A measure of vector boson production at the LHC may serve as a starting point for further experimental explorations of TMDs.
	\item These results may provide an understanding of the physics in the saturation region, where the density of partons is very high and the production of partons and their  increasing  at low x, occur at a constant rate. In this region, the transverse momentum distributions of the sea quarks and gluons can be considered as the Fractal or self-similar distributions with constant dimensions, which means that when you change x, for example, the behavior of the TMDs does not change, and then they can be described by the same exponents. Gluons must be produced and decayed at constant rates, and the parton densities must increase with a constant gradient.  This issue requires further investigation, of course.
\end{itemize}

%\clearpage

%
%%%%%%%%%%%%%%%%%%%%%%%%%%%%%%%%%%%%%%%%%%%%%%%%%%%%%%%%%%%%%%%%%%%%%%%
\section{ Summary and Conclusions} \label{sec:conclusion}
%%%%%%%%%%%%%%%%%%%%%%%%%%%%%%%%%%%%%%%%%%%%%%%%%%%%%%%%%%%%%%%%%%%%%%%
High-precision data has been collected by the HERA experiment on electron (positron)-proton deep inelastic scattering processes. This data covers the low-$x$ region where partons (sea quarks and gluons) carrying a very small fraction of the proton's momentum become abundant. In this work, we aim to describe these parton distribution functions (PDFs) at low-$x$ using a Fractal-based approach.
	Our analysis is performed at the next-to-leading order (NLO) in QCD perturbation theory. It is demonstrated that parameterizing the PDFs based on Fractal distributions allows us to phenomenologically model them below $x<0.01$ and describe the experimental HERA data well. Another advantage of this parameterization is its ability to reproduce neutral current interactions in the combined HERA data set down to $x<0.01$ and low $Q^2$ values of up to 1.2 GeV$^2$.
	To assess the uncertainties in the resulting low-$x$ PDFs and their impact on observables, we have extensively employed the Hessian method. This uncertainty analysis provides insights into how measurement uncertainties propagate and affect our understanding of the parton content of the proton at small momentum fractions, where nonlinear QCD effects become important. Continued improvements in precision and kinematic reach, for example from LHeC, could help further constrain Fractal-inspired PDF models and test predictions regarding the onset of parton saturation and other emergent phenomena. Overall, our study demonstrates the potential of Fractal-based approaches to successfully parametrize nonperturbative aspects of QCD dynamics. In this paper, we show that the geometric scaling behavior observed in HERA data supports the idea of self-similar Fractal distributions. Precise structure function measurements may be used to quantitatively determine Fractal scaling exponents with more accuracy.

%
%%%%%%%%%%%%%%%%%%%%%%%%%%%%%%%%%%%%%%%%%%%%%%%%%%%%%%%%%%%%%%%%%%%%%%%
\section*{Acknowledgments}
%%%%%%%%%%%%%%%%%%%%%%%%%%%%%%%%%%%%%%%%%%%%%%%%%%%%%%%%%%%%%%%%%%%%%%%
%
We gratefully acknowledge to Professor F. Olness, Dr. Sara Taheri Monfared and Dr. Hamed Abdolmaleki for carefully reading the manuscript, many
helpful discussions, and comments. S. A. T. is grateful to the School of Particles and Accelerators, Institute for Research in Fundamental
Sciences (IPM) to make the required facilities to do this project.  We are thankful to Ferdowsi University of
Mashhad for the financial support provided for this research. This work is supported by the Ferdowsi
University of Mashhad under grant number 2/53356(1399/10/09).

\end{document}